\documentclass[a4paper,10pt, draftcls, onecolumn]{IEEEtran}
\usepackage{amssymb,amsmath}
\usepackage{amsmath}
\usepackage{threeparttable}   
\usepackage[dvips]{color}
\usepackage{url}   
\usepackage[T1]{fontenc}
\usepackage{algorithm}               
\usepackage{algorithmic}    
\usepackage{cite}
\usepackage{multirow}
\usepackage[dvips]{color}
\usepackage{lineno}
\usepackage[normalem]{ulem}
\usepackage{color,soul}  
\usepackage{indentfirst}
\usepackage{setspace}
\usepackage{cases}
\usepackage{tikz}
\usepackage{algorithm}
\usepackage{algorithmic}

\begin{document}
\title{Stochastic Channel Models for Massive and XL-MIMO Systems}
\author{L\'igia May Taniguchi, Taufik Abr\~ao
\thanks{This work was supported in part by CAPES, Financial Code 001, the  Arrangement between the European Commission (ERC) and the Brazilian National Council of State Funding Agencies (CONFAP), CONFAP-ERC Agreement H2020, by the National Council for Scientific and Technological Development (CNPq) of Brazil under grant 310681/2019-7.}
\thanks{L. M. Taniguchi and T. Abr\~{a}o are with Electrical Engineering Department, State University of Londrina (UEL), Londrina, PR, Brazil;  \texttt{ligia.lmt@gmail.com;  taufik@uel.br}.}}

\maketitle

\begin{abstract}
In this paper, {stochastic channel models for massive MIMO (M-MIMO) and extreme large MIMO (XL-MIMO) system applications are described, evaluated and systematically compared.} This work aims to cover new aspects of massive MIMO stochastic channel models in a comprehensive and systematic way. For that, we compare  different models, presenting graphically and intuitively the behavior of each model. Each massive MIMO channel model emulates the environment using different methodologies and properties. Using metrics such as capacity, SINR, singular values decomposition (SVD), and condition number, one can understand the influence of each characteristic on the modelling and how it differentiates from other models. Moreover, {in new XL-MIMO scenarios, where the near-field and visible region (VR) effects arise,} our finding demonstrate that for the two assumed schemes of clusters distribution, the clusters location influences the performance of the conjugate beamforming and zero-forcing (ZF) precoding due to the correlation effect, {which have been analysed from the geometric massive MIMO channel models}.
\end{abstract}


\section{Introduction}\label{sec1}

In the last decades, wireless communications have become  increasingly indispensable, so numerous applications have been created, improving the demand of capacity and reliability of wireless system \cite{Lu2014}. One way to develop such requirements consists in applying the {Massive Multiple-input-multiple output} (M-MIMO), considered a key technology, {in which the Base Station (BS) is equipped with a large number of antennas \cite{Larsson2014}. In M-MIMO, the number of BS antennas $M$ is typically of order of hundreds, being limited in \cite{bjrnson2019massive} by $M \geq 64$.} However, this technology presents many challenges as the propagation channel modeling. The channel represents a fundamental part in wireless communications, responsible for causing {strong degradation in the received signal, and as consequence a remarkable variation in the decoded signal and the overall system performance.}  Thus, a realistic propagation model is needed to understand how the environment can change the signal {under large as well extra-large BS antenna arrays and mobile configurations of the terminals.} Although we have many channel models describing MIMO scenario, unfortunately, there is still no standard channel model available to completely describe all the channel characteristics of the M-MIMO {and extreme large massive MIMO (XL-MIMO) system} scenarios. The fact of M-MIMO has many tens of antennas, typically hundreds of antennas at BS, impacts on the unique characteristics, {of such MIMO systems, including} favorable propagation and hardening channel \cite{marzetta_book2016}, elevation characteristics due to 2D or 3D antennas array, spherical wave-front assumption \cite{Tamaddondar2017} and spatial non-stationarity \cite{carvalho2019nonstationarities}, {the last two features are a consequence of the near-field propagation waveform, which occurs typically in XL-MIMO configurations.}.   

The favorable propagation {offers} a mutual orthogonality between channel vectors {from different mobile terminals (MTs);} hence, the application of linear signal processing techniques at the receiver side can result in  optimal performance \cite{marzetta_book2016, Bjornson2016}. The antenna elements may be arranged in different structures and generate a radiation pattern according to its structure \cite{Vesa2015}. The majority of the works consider a uniform linear  array (ULA) structure; however, only 2D or 3D structures such as uniform planar array (UPA) and uniform cylindrical array (UCA) offer control of angle of elevation, resulting in an increase of spatial resolution, \textit{i.e.}, an increasing on the desired signal strength {while simultaneously provide} reduction in users' interference \cite{Zheng2014}.  

{In the literature, there are several channel models that strive to match the spatial correlation in M-MIMO channels, the classical exponential correlation model being one of these.  Authors of \cite{Croisfelt2019} analyze how the channel estimation is affected by the correlated fading model; for that, they investigate an M-MIMO scenario applying the standard MMSE channel estimation approach over uniform linear and planar arrays (ULAs and UPAs, respectively) of antennas. The spatially correlated channels generated by this combined model results in an improved channel estimation quality. The UPA acquired better results regarding pilot contamination since it has been demonstrated that this type of array generates stronger levels of spatial correlation w.r.t. the ULA. In contrast to the advantageous results in channel estimation, the channel hardening effect was impaired by the spatially correlated channels, with higher system performance degradation equipped with UPA.}

From the measurements, wave-fronts should be assumed to be spherical {for typical XL-MIMO scenarios.} The spherical wave channel modeling based on electromagnetic field describes with more accuracy the near-field propagation if compared to plane-wave propagation; hence the models present different results in capacity \cite{Miao2018, Tamaddondar2017}. Moreover, the non-stationary properties of the channel clusters can be observed over the large antenna arrays due to the possibility of different antenna elements observe different sets of clusters \cite{Chen2017}. This property is paramount in scenarios XL-MIMO and must be considered from a spatial perspective.

In the literature, there are two approaches for wireless massive MIMO channels modeling:  deterministic and statistical. The first approach is based on electromagnetism theory and is considered more accurate{; generally the deterministic channel models  are described considering physical aspects and the geometry of the propagation channel}. However, the computational burden makes the model unworkable in practical terms. Although the statistical channel models result less accurate, they present reduced computational complexity, making expedite their implementation and analysis. 

Among the {statistical M-MIMO channel models}, there are two kinds of channel models, namely, \textit{correlation-based stochastic models} (CBSM) and \textit{geometry-based stochastic models} (GBSM). The GBSM is categorized in 2D and 3D according to the antennas array used. Such models are usually accurate and flexible for describing different scenarios, {because the model can consider the environmental characteristics as, for example, the type of distribution of the scatterers and the locations where the scatterers are distributed.} In GBSM, such characteristics result in a correlation degree between the antennas signals; however, in CBSM, this correlation degree is only modeled by a correlation factor, making the CBSM model less complex than the {geometric-based models.} 

Recently, a very new scenario has being studied to be implemented in {5G and beyond communication systems, deploying an extremely large number of antennas at the BS in one, two or three-dimensions antenna array arrangements, namely XL-MIMO systems}. The main idea consists in distribute the antenna elements, for instance, on {the entire} wall of buildings, creating antenna array {placement in order of thousands antenna elements.} Therefore, unlike the previous models available in the literature, {the wireless channel models should cover new physical characteristics, including} non-stationarities and near-field propagation, impacting on the construction and usability of the wireless channel models \cite{carvalho2019nonstationarities}.

{Very recent researches on the non-stationarities channel modeling include \cite{Chen2017, Jiang18, Wu2015, ali2019,Han2020}. In \cite{Jiang18} a channel model was proposed for vehicle-to-vehicle communication environments, in which the spherical wave-front is assumed. Also, the non-stationarity property is evaluated, however, in a temporal sense. In \cite{Chen2017, Wu2015}, the spherical wave-front and the spatial non-stationarity property were evaluated, considering the non-stationarity properties on the array for M-MIMO channel modeling. For XL-MIMO systems, in \cite{ali2019}, linear receivers are evaluated, where the non-stationarities are included in the correlation matrix from each user. In \cite{Han2020}, the near-field propagation considering the spherical wave-front was considered, including the spatial non-stationarity properties for XL-MIMO systems. However, the methodology to channel modeling considered only LoS paths even with the presence of scatterers.}

{ The work contributions are fourfold. First, the paper covers new aspects of massive XL-MIMO stochastic channel modeling in a comprehensive and systematic way, with focus on the comparison and performance evaluation of various existing channel models, novel cluster distribution scheme and visibility regions (VRs) generation for XL-MIMO systems. Second, using figure of merit such as capacity, singular values decomposition (SVD) and condition number (CN), we discuss the influence of each characteristic on the stationary models and how it differentiates from other models. Third, based on a physical perspective, we propose an algorithm to generate  the VRs in the new XL-MIMO scenarios, deploying the SINR metric for the analysis, considering two classical linear precoders. The algorithm considers realistic massive MIMOchannel configurations, including the presence of obstacles between the scatterers and the antennas array, being also considered the size of the clusters which defines the VR length along the elements of antenna array. Finally, we provide and analyze extensive numerical results to characterize the correlation-based and geometric-based stochastic channel models for massive and XL-MIMO equipped with uniform linear and planar arrays subject to spatial correlation, and also taking into account the large-scale fading component. }

 The rest of the paper is organized as follows.  Section \ref{sec:channel} describes the stochastic channel models for {massive and extreme large} MIMO systems. {The main figures of merit deployed in the caracterization of geometric and correlation-based channels are described in section \ref{sec:FoM}, while} section \ref{sec:results} evaluates the stochastic channel models in terms of achievable capacity, SINR, condition number and SVD analysis. {The main} conclusions are offered in section \ref{sec:concl}.\\

\section{Channel Models for M-MIMO and XL-MIMO}\label{sec:channel}
\begin{figure}[!htbp]
\centering
\includegraphics[width=0.75\textwidth]{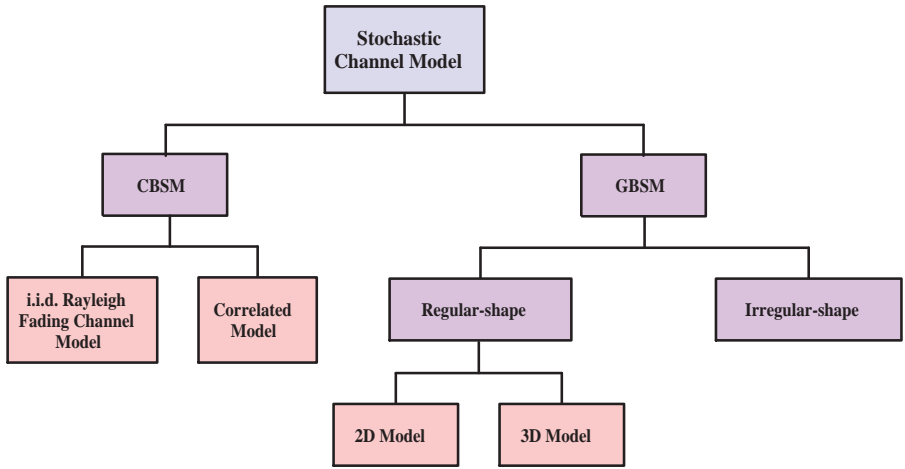}
\caption{Stochastic Channel Models.}
\label{fig:1}
\end{figure}

The stochastic channel models are defined in CBSM and GBSM as illustrate in Fig. \ref{fig:1}. The CBSMs can be categorized into two types known as classic i.i.d. Rayleigh fading channel model and correlated channel models \cite{Wang2016} that describe the characteristics by correlation matrices. To GBSM, works usually assume the one-ring, two-ring and elliptical-ring scenarios \cite{Yu2002,Bakhshi2008,Arias2002}. These models are frequently implemented in scenarios where the BS can employ hundreds of antennas in compact arrays. However, a recent scenario, called XL-MIMO, was proposed to improve the area throughput in wireless networks. In this scenario, the antennas elements are disposed in a large surface,  {where the number of antenna elements is of order of five hundreds or even thousands \cite{bjrnson2019massive} antenna elements}. Next, we describe the propagation channel models considered.

{In addition, channel models can be classified regarding the dimensions occupied by the antenna elements arrangements, {\it i.e.,}  1D, 2D or 3D antenna elements arrangements; such arrangements are associated typically to uniform linear, planar and cylindrical antenna array structures (ULA, UPA and UCA, respectively). Such classification is useful since the geometry of arrangement, as well as the number of antenna-elements  can be deployed to control the spatial diversity and the direction of signal propagation;  hence,  the linear arrangement is able to control the signal propagation only changing the azimuth angles; while planar, spherical, cylindrical array shapes explore the spatial diversity controlling simultaneously azimuth and elevation angles. 2D-GMSM and 3D-GBSM are described in subsection \ref{sec:2d-gbsm} and \ref{sec:3d-gbsm}, respectively.  For instance, the 3D-GBSM models combine geometry-based stochastic channel models with 3D antenna array arrangements, being able to control the radiation and reception of signals to any direction in 3D space, as defined by azimuth and elevation angles. In practical communication systems, 2D and 3D antenna arrangement models are more likely to be more widely used \cite{Zheng2014} due to the attainable higher spatial diversity}. 

\subsection{CBSM}

In certain situations, a low-complexity and mathematically tractable massive MIMO channel model is preferred when analysing and simulating system performance \cite{Wang2016}. Thus, many researches work with {MIMO channel models based on spatial antenna correlation.} The first model is the classical exponential model of channel correlation matrix described in the following.  Next, we analyze the uncorrelated as well the exponential with large-scale fading MIMO channel models.

\subsubsection{Exponential Spatial Correlation}
The first channel model is the classical exponential model \cite{Loyka2001}, {represented by a Toeplitz channel spatial antenna correlation matrix}: 
\begin{equation}
[\textbf{R}]_{m,n}= 
\begin{cases}
    \rho^{n-m},& \text{if } m\leq n\\
     \rho_{nm}^*,    & \text{otherwise}, \qquad n,m=1, 2, \ldots M
\end{cases}
\label{eq:1}
\end{equation}
\noindent where {$M$ is the total number of antennas uniformly arranged in rows and columns planar or linear structure (UPA or ULA);}  $(mn,)$ is the $mn$-th element of the correlation matrix that correspond to the {position of antenna element in the planar antenna array structure, while $\rho\in[0;\, 1]$ is the correlation factor, which value depends on how close the antenna-elements are each other.}

\subsubsection{Uncorrelated Fading with Large-Scale Fading}
A consideration of a uncorrelated model is common in the literature {by setting the correlation matrix in \eqref{eq:1}} as $\textbf{R}=\beta \textbf{I}_{M}$, where $\beta$ is the path loss term defined as squared amplitude. Adding the effects of the shadowing \cite{Emil2018}, the uncorrelated channel model is described as:
\begin{equation}
\textbf{R}=\beta \hspace*{0.05cm}\textnormal{diag}(10^{\frac{f_1}{10}}, \dots , 10^{\frac{f_M}{10}})
\label{eq:4}
\end{equation}
\noindent where $f_1, \dots,f_M  \sim \mathcal{N}(0,\,\sigma_{\text{shad}})$ are the random fluctuation of the large-scale fading.

\subsubsection{Exponential Spatial Correlation with Large-Scale Fading}
A model discussed in \cite{Emil2018} and based on \cite{Loyka2001} combines \eqref{eq:1} and \eqref{eq:4}, resulting in an exponential spatial correlation with the presence of shadowing effects:
\begin{equation}
[\textbf{R}]_{m,n}=\beta \cdot \rho^{|n-m|}\cdot e^{i(n-m)\theta} \cdot 10^{\frac{f_m+f_n}{20}}
\label{eq:5}
\end{equation}
\noindent where $\theta$ is the Angle-of-Arrive (AoA). This latter model is more complete as it encompasses more features than both \eqref{eq:1} and \eqref{eq:4} exponential-based channel models.

\subsection{2D GBSM}\label{sec:2d-gbsm}

{GBSM can be classified according to the distribution of the scatterers combined with the number of dimensions of the antenna elements placement. When the propagation environment is analyzed, the GBSM can be classified in {\it regular-shape} (RS-GBSM) and {\it irregular-shape} (IR-GBSM) geometry-based stochastic models. In this case, the scatterers can be distributed on regular shapes (for example one-ring, two-ring and ellipse) or irregularly (randomly distributed) \cite{Yin2016}.} 

The One-ring channel model is appropriate for describing environments, in which the base station is elevated and unobstructed, whereas the user equipment (UE) is surrounded by a large number of local scatterers, while the two-ring and elliptical model is appropriate for environments in which both base station and the users are surrounded by local scatterers \cite{Ptzold2012}. Although these models are well established in the literature, such models do not accurately describe {M-MIMO and XL-MIMO configurations and scenarios}. Although currently there is no model that describes all the characteristics present in the XL-MIMO environment, methodologies that partially describe such scenario have been elaborated, {including the effects of the} non-stationary and spherical wave-front characteristics.

{In this paper, we use the 2D and 3D RS-GBSMs classification, considering the One-ring and Gaussian Local Scattering to define the distribution of the scatterers. In One-ring scenarios, the scatterers are uniformly spread around the user while for a scenario called Gaussian Local Scattering, similar to One-ring, the location of the scatterers present Gaussian distribution.} For the 2D-GBSM, we consider that the antennas are linearly arranged, also known as ULA, where the array response is given by \cite{massivemimobook}:
\begin{equation}
\textbf{a}_n = g_n [1 \hspace{0.1cm} e^{2\pi i d_H sin(\overline{\varphi}_n)} \dots  e^{2\pi i d_H (M-1)sin(\overline{\varphi}_n)}] ^T
\label{eq:6}
\end{equation}
\noindent where $g_n \in \mathbb{C}^M$ is the average gain of the n-th multipath component, the $\overline{\varphi}$ is the angle of an arbitrary multipath component and $d_H$ is the antenna spacing  in wavelength of the array . Thus, the channel response $\textbf{h}$ is the superposition of the array responses of the $N_{path}$ components as show:
\begin{equation}
\textbf{h} = \sum_{n=1}^{N_{path}} \textbf{a}_n 
\label{eq:7}
\end{equation}

Hence, we have the correlation matrix of the channel presented in eq. (\ref{eq:7}) as:
\begin{equation}
\textbf{R}=\mathbb{E} \Bigg \{\sum_{n=1}^{N_{\rm path}} \textbf{a}_n \textbf{a}_n^H \Bigg \}
\label{eq:8}
\end{equation}
\noindent where we can rewrite the equation above for the (m,n)th element of \textbf{R} as:
\begin{equation}
[\textbf{R}]_{m,n}=\sum_{n=1}^{N_{\rm path}} \mathbb{E}\{|g_n|^2\} \mathbb{E}\{e^{2 \pi i d_H (m-1) sin(\overline{\varphi}_n)} e^{-2 \pi i d_H (n-1) sin(\overline{\varphi}_n)}\}
\label{eq:9}
\end{equation}

Considering the total average gain of the multipath components $ \mathbb{E}\{|g_n|^2\}$ as $\beta$ and applying the expectation operator, we have:
\begin{equation}
[\textbf{R}]_{m,n}=\beta \int_{-\infty}^{\infty}  e^{2 \pi i d_H (m-n) sin(\overline{\varphi})} f(\overline{\varphi}) d\overline{\varphi}
\label{eq:10}
\end{equation}
\noindent where $ f(\overline{\varphi})$ is the PDF of $ \overline{\varphi}$.

\begin{figure}[!htbp]
\centering
\includegraphics[width=0.75\textwidth]{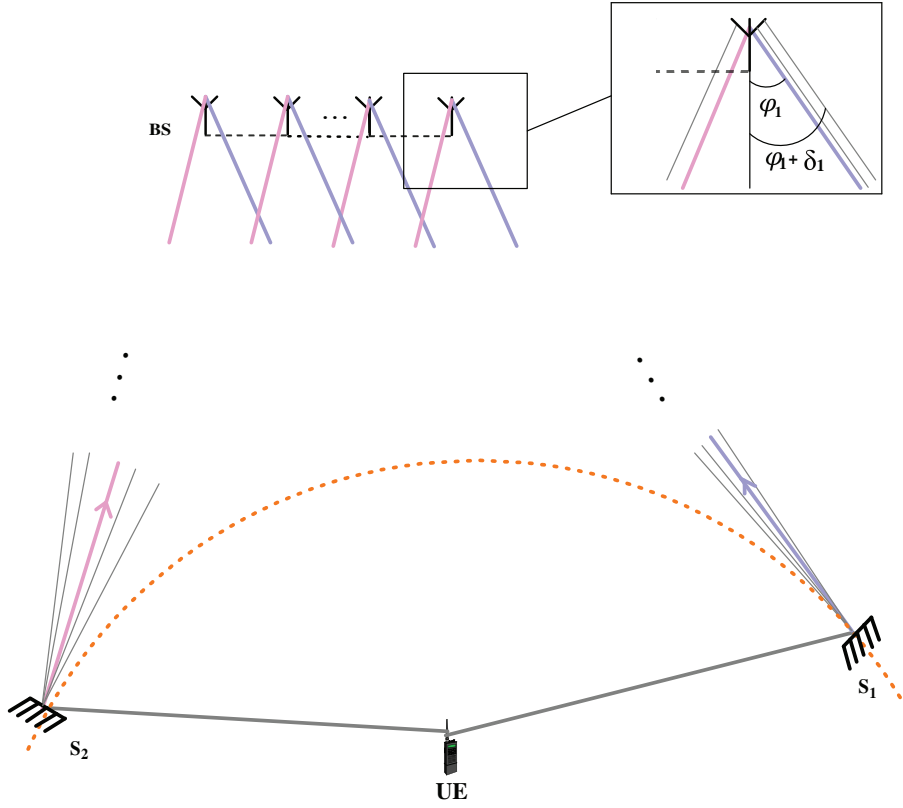}
\caption{Geometry-based Stochastic Model in Uplink mode where the BS is equipped with linear array. The BS and user are far, thus, the wave-front can be approximated to a plane.}
\label{fig:16}
\end{figure}

The model in eq. (\ref{eq:10}) is a generic model to GBSM which $f(\overline{\varphi})$ can assume uniform, gaussian and laplace distribution. Illustratively, we describe geometrics models as in Fig. \ref{fig:16}, where multiple signals arrives in all antennas. In the figure, $\varphi_1$ is the nominal angle and $\delta_1$ is the angular standard deviation derivated of mutipath signals, being both originated from the scatterer $S_1$. 

\subsubsection{One-ring Model} 
If we consider $ f(\overline{\varphi})$ as a uniform distribution, we have the one-ring model which will be presented below. The one-ring model is defined as a local scattering, around to UE, distributed uniformly as $f(\overline{\varphi})   \sim \mathcal{U} (\varphi-\Delta,\varphi+\Delta)$. Thus, the correlation matrix will result in:
\begin{equation}
[\textbf{R}]_{m,n}=\frac{\beta}{2\Delta}\int_{\varphi-\Delta}^{\varphi+\Delta}  e^{2 \pi i d_H (m-n) sin(\overline{\varphi})} d\overline{\varphi}
\label{eq:11}
\end{equation}

Considering that $\overline{\varphi}=\varphi+\delta$ the equation can be rewritten as:  
\begin{equation}
[\textbf{R}]_{m,n}=\frac{\beta}{2\Delta}\int_{-\Delta}^{\Delta}  e^{2 \pi i d_H (m-n) sin(\varphi+\delta)} d\delta
\label{eq:12}
\end{equation}
\noindent where $\varphi$ is the nominal angle of arrival (AoA) and $\delta$ is the variation around to nominal angle, $\Delta$ is the range in which the multi components are disposed. {Such variables, AoA and $\Delta$, represent the degree of channel correlation. Thus, the correlation between the signals can be modeled by the angular interval at which the signals arrive while the degree of correlation between the antennas can be modeled from AoA.}

\subsubsection{Gaussian Local Scattering Model} 
Another channel model based in geometry is the Gaussian local scattering model described in \cite{massivemimobook}, where the correlation matrix is described by:
\begin{equation}
[\textbf{R}]_{m,n}=\beta \int_{-\infty}^{\infty}  e^{2 \pi i d_H (m-n) sin(\overline{\varphi})} \frac{1}{\sqrt{2 \pi}\sigma_{\overline{\varphi}}}e^{- \frac{(\overline{\varphi}-\varphi)^2}{2 \sigma_{\overline{\varphi}}^2}} d\overline{\varphi}
\label{eq:17}
\end{equation}

Taking the same consideration as the one-ring model, $\overline{\varphi}=\varphi+\delta$, the eq. in (\ref{eq:17}) can be rewritten as:
\begin{equation}
[\textbf{R}]_{m,n}=\beta \int_{-\infty}^{\infty}  e^{2 \pi i d_H (m-n) sin(\varphi+\delta)} \frac{1}{\sqrt{2 \pi}\sigma_{\varphi}}e^{- \frac{\delta^2}{2 \sigma_\varphi^2}} d\delta
\label{eq:16}
\end{equation}
\noindent where $\sigma_\varphi$ is the angular standard deviation (ASD) of the multiple signal components around the nominal angle. An approximation can be done if $\sigma_{\varphi}$ has small values, below $15^{\circ}$ resulting in a close-form expression:
\begin{equation}
[\textbf{R}]_{m,n}=\beta e^{2 \pi i d_H (m-n) sin(\varphi)} e^{-\frac{\sigma_\varphi^2}{2} (2 \pi d_H (m-n) cos(\varphi))^2}
\label{eq:13}
\end{equation}

Some channel measurements showed the presence of shadowing \cite{gao2015} and in \cite{Sanguinetti2019} the correlation matrix for the gaussian model was presented as:
\begin{equation}
[\textbf{R}]_{m,n}=\beta 10^{\frac{f_{m}+f_{n}}{10}}  \frac{1}{S}\sum_{s=1}^{S} e^{2 \pi i d_H (m-n) sin(\varphi_s)} e^{-\frac{\sigma_\varphi^2}{2} (2 \pi d_H (m-n) cos(\varphi_s))^2}
\label{eq:15}
\end{equation}

\noindent where $S$ is the number of scatterers and $\varphi_s$ is the nominal angle of arrival of the s-th scatterer. 

\subsection{3D GBSM} \label{sec:3d-gbsm}
The 3D GBSMs are structure as planar, cylindrical and spherical capable of creating beams controled by two angles, azimuth and elevation \cite{Zheng2014}. Most of the channel model described for the MIMO system considers 2D models, however, for applications on M-MIMO systems, the 3D models are better candidates due to the large number of antennas that M-MIMO proposes \cite{Xie2015}. {Based on \cite{massivemimobook}}, a planar strutucture, known as Uniform Planar Array (UPA), will be analyzed with the One-ring model and the Gaussian Local Scattering model.

When we consider that the user is far enough from the antennas array, the plane wave can be evaluated as:
\begin{equation}
\textbf{k}(\overline{\varphi},\overline{\theta})=\frac{2\pi}{\lambda}
\begin{bmatrix} 
\cos(\overline{\theta})\cos(\overline{\varphi}) \\
\cos(\overline{\theta})\sin(\overline{\varphi}) \\
\sin(\overline{\theta})
\end{bmatrix}
\label{eq:28}
\end{equation}

\noindent where $\overline{\varphi}$ and $\overline{\theta}$ are the azimuth and elevation angles of each received signal. The location of the m-th antenna on the x, y, and z axis for the planar array can be described as:
\begin{equation}
\textbf{u}_m = 
\begin{bmatrix} 
0 \\
p_y(m)d_H \lambda \\
p_z(m)d_V \lambda
\end{bmatrix}
\label{eq:29}
\end{equation}

\noindent where $p_y(m) = \rm mod(m-1, M_H)$ and $p_z(m) = \lfloor (m-1)/M_H \rfloor$ are the horizontal and vertical index of antenna $m$, respectively. Thus, we can write the correlation matrix as:
\begin{equation}
[\textbf{R}]_{m,n}= \beta \int \int e^{j 2\pi d_V[p_z(m)-p_z(n)]sin(\overline{\theta})} e^{j2\pi d_H[p_y(m)-p_y(n)]cos(\overline{\theta})sin(\overline{\varphi})} f(\overline{\varphi},\overline{\theta}) d\overline{\varphi} d\overline{\theta}
\label{eq:33}
\end{equation}

\subsubsection{One-ring Model} 
From eq. (\ref{eq:33}), if we consider that elevation and azimuth angles with uniform distribution, we have a 3D one-ring. Thus, the one-ring model for a UPA is decribed as:
\begin{equation}
[\textbf{R}]_{m,n}=\frac{\beta}{4\Delta_{\varphi}\Delta_\theta} \int_{\theta - \Delta_{\theta}}^{\theta + \Delta_{\theta}}\int_{\varphi - \Delta_{\varphi}}^{\varphi + \Delta_{\varphi}} e^{i2\pi d_V[p_z(m)-p_z(n)]sin(\overline{\theta})} e^{i2\pi d_H[p_y(m)-p_y(n)]cos(\overline{\theta})sin(\overline{\varphi})} d\overline{\varphi} d\overline{\theta}
\label{eq:27}
\end{equation}
\noindent where $\overline{\theta} \in [-\pi/2,\pi/2)$ and $\overline{\varphi} \in [- \pi,\pi)$, $\Delta_{\theta}$ and $\Delta_{\varphi}$ are the angular spread in elevation and azimuth domain,  $p_z(m)$ and $p_y(m)$ refers to the index of the m-th antenna on the z and y axis, respectively, in the same sense, $p_z(n)$ and $p_y(n)$ refers to the index of the n-th antenna on the z and y axis. The horizontal and vertical distances are defined by $d_H$ and $d_V$ with the antenna element placed by $x$ (row index) and $y$ (column index) and $\beta$ describes the macroscopic large-scale fading.

Considering $\overline{\varphi} = \varphi + \delta_{\varphi}$ and $\overline{\theta} = \theta + \delta_{\theta}$, the eq. (\ref{eq:27}) is rewritten as:
\begin{equation}
[\textbf{R}]_{m,n}=\frac{\beta}{4\Delta_{\varphi}\Delta_\theta} \int_{- \Delta_{\theta}}^{\Delta_{\theta}}\int_{- \Delta_{\varphi}}^{\Delta_{\varphi}} e^{i2\pi d_V[p_z(m)-p_z(n)]sin(\theta + \delta_{\theta})} e^{i2\pi d_H[p_y(m)-p_y(n)]cos(\theta +\delta_{\theta}) sin(\varphi + \delta_{\varphi})} d\delta_{\varphi} d\delta_{\theta}
\label{eq:36}
\end{equation}

{In the same way of 2D models, in 3D models both azimuth and elevation angles play a fundamental role in channel modeling, incorporating information that defines the degree of correlation of the antennas. In addition, correlation can also be modeled by the angular interval, $\Delta_{\varphi}$ and $\Delta_{\theta}$, in which it defines the correlation between the received signal replicas, imposed by the environment (via scatters).}

\subsubsection{Gaussian Local Scattering Model} 
From eq. (\ref{eq:33}), we will analyze the scattering with Gaussian distribution shown as:
\begin{equation}
[\textbf{R}]_{m,n}=\frac{\beta}{2\pi \sigma_{\varphi} \sigma_{\theta}} \int_{- \infty}^{\infty}\int_{- \infty}^{\infty} e^{i2\pi d_V[p_z(m)-p_z(n)]sin(\theta)} e^{i2\pi d_H[p_y(m)-p_y(n)]cos(\theta)sin(\varphi)}  e^{\frac{1}{2 \sigma_{\varphi}^2 }(\overline{\varphi}- \varphi)^2 } e^{\frac{1}{2 \sigma_{\theta}^2 } (\overline{\theta} - \theta)^2)} d\overline{\varphi} d\overline{\theta}
\label{eq:26}
\end{equation}
\noindent where $\sigma_{\varphi}$ and $\sigma_{\theta}$ describe the angular standard deviation of the azimuth and elevation angles. Considering $\overline{\varphi} = \varphi + \delta_\varphi$ and $\overline{\theta} = \theta + \delta_\theta$, the eq. (\ref{eq:26}) is rewritten as:
\begin{equation}
[\textbf{R}]_{m,n}=\frac{\beta}{2\pi \sigma_{\varphi} \sigma_{\theta}} \int_{- \infty}^{\infty}\int_{- \infty}^{\infty} e^{i2\pi d_V[p_z(m)-p_z(n)]sin(\theta)} e^{i2\pi d_H[p_y(m)-p_y(n)]cos(\theta)sin(\varphi)}    e^{\frac{\delta_\varphi^2}{2 \sigma_{\varphi}^2 } + \frac{\delta_\theta^2}{2 \sigma_{\theta}^2}} d\delta_\varphi d\delta_\theta
\label{eq:29}
\end{equation}

\subsection{Extreme Large Massive MIMO Channels}
Massive MIMO is a system in which the BS is equipped with hundreds of antennas. The large number of antennas impacts in some unique characteristics, such as spherical wave-front  (near-field) assumption and spatial non-stationarity. In MIMO system, the assumption of the plane wave-front is reasonable since the size of antennas array is smaller, \textit{i.e.}, the power variations between the antennas can be assumed approximately a constant. However, in XL-MIMO the number of antennas and the array dimension can be considered extreme large, while the users distance to the array is very short compared to the array size. Thus, the far-field assumption does not hold, being necessary the spherical wave-front assumption. The far-field is often taken to exist for distances greater than the Rayleigh distance, given as: 
\begin{equation}
Z=\frac{2D^2}{\lambda}
\label{eq:38}
\end{equation}
\noindent where $D$ is the maximum dimension of the antenna or antenna array, and $\lambda$ represents the wavelength.

\begin{figure}[!htbp]
\centering
\includegraphics[width=.9\textwidth]{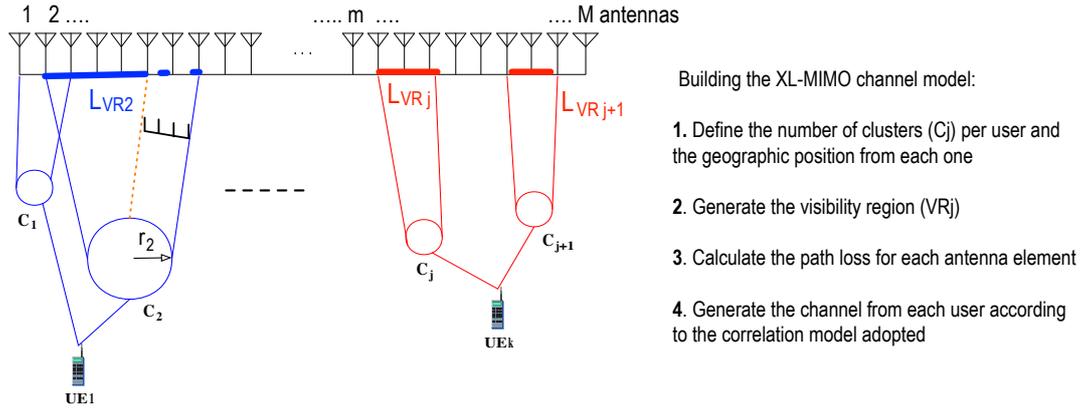}
\vspace{-2mm}
\caption{Typical XL-MIMO scenario: each cluster defines the VR length ($L_{\textsc{vr}_j}$) in the antennas array.}
\label{fig:51}
\end{figure}

The non-stationary properties associated to the channel clusters can be observed on large antenna arrays due to the possibility of different antenna elements observe different sets of clusters \cite{Chen2017}, where the set of antenna elements observed is called visibility region (VR).  In massive MIMO configurations, it is observed significant non-stationaries across antennas elements when the mobile terminals are close to the antenna array, as reported by measurements in \cite{carvalho2019nonstationarities}, being necessary consider from a spatial perspective \cite{oestges2007}. The XL-MIMO principle consists in incorporating the elements of antennas in a building of large dimension, thus, the non-stationarities and the spherical wave-front features become essential considerations. The physical  XL-MIMO channel scenarios with a linear array and $C_j$ channel clusters can be represented as in Fig. \ref{fig:51}.
To build the XL-MIMO channel model, one can follow the steps as determine the number of cluster and the geographical position;  generate the VR for each cluster;  determine the pathloss for each antenna element; and generate the channel coefficient from each user according to the antenna correlation model.

\begin{figure}[!htbp]
\centering
\includegraphics[width=.65\textwidth]{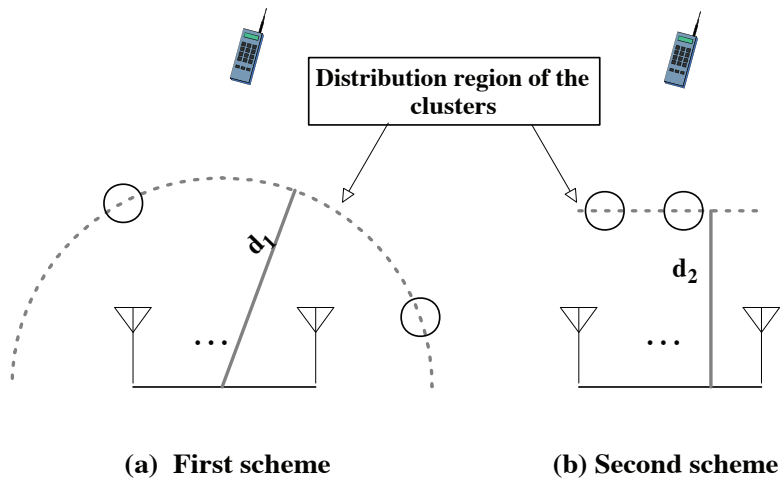}
\vspace{-2mm}
\caption{Adopted channel clustering distribution following two representative scenarios.}
\label{fig:41}
\end{figure}

From the perspective of clustering position, in this work we analized two schemes scenarios, as depicted in Fig. \ref{fig:41}. In the first configuration, a distance $d_1$ corresponding to an equal distance between the center of the BS and any cluster positioning in the valid distribution cluster region. Thus, the $d_1$ and azimuth angle will determine the cluster localization, as well as the received power in each BS antenna array. In the second scheme, the cluster distribution region is parallel to the BS and the distance between the region and the array is determined by $d_2$.

In step 2 of Fig. \ref{fig:51}, we need to define the size of each cluster, because it will determine the VR size. Hence, in Fig. \ref{fig:51} one can see that the cluster $C_2$ is able to see a larger number of antennas compared to cluster $C_1$ due to the cluster size. In this work, we define  the cluster radius $r$ and the VR size region $L_{\text{VR}}$ are related by $L_{\text{VR}}=2 r$. Moreover, in the VR it is possible the presence of obstacles, reducing the number of antennas that receive the signal from the corresponding channel cluster. Hence, considering the number of antennas that the cluster can observe, a simple way to simulate the obstacles consists in generate the VR in XL-MIMO scenarios as presented in Algorithm \ref{algo:VR}, where $\lambda$ is the carrier wavelength,  $p_0$ is the probability of the antenna not being visible, $p_1$ is the probability of the antenna being visible, $c$ is a factor which defines how fast the regions changes to visible and not visible (amount of obstacles), and $r_\text{min}$ and $r_\text{max}$ are the minimum and maximum cluster radius, respectively.

In step 3, Fig. \ref{fig:51}, the path loss in dB from the user to each antenna element $n$ is calculated as:
\begin{equation}
\mathcal{L}(n)=\mathcal{L}_0 - 10 \cdot \alpha \cdot  \log_{10}\left[\frac{d(n)}{d_0}\right]
\label{eq:40}
\end{equation}
\noindent where $\mathcal{L}_0$ is the channel average gain at the reference distance of $d_0=1$m, $\alpha$ is the path loss exponent, which determines how fast the signal power decays with the distance, $d(n)$, in meters, is the sum of the distance between the $c$th cluster to the $n$th antenna element and the distance between the $k$th UE to the $c$th cluster.

Finally, in the step 4 the channel from each cluster $c$ corresponding to each user $k$ is generated as:
\begin{equation}
\textbf{h}_{k,c}=\boldsymbol{\beta}_{k,c} \circ \,  (\textbf{R}_{k,c}^{\frac{1}{2}} \textbf{z}_{k,c})
\end{equation}
\noindent where $\circ$ denotes the Hadarmard product, $\boldsymbol{\beta}_{k,c}$ is the vector of path loss amplitude gain, related to user $k$ and cluster $c$, while $\textbf{R}_{k,c}$ is the correlation matrix for the cluster $c$ and user $k$, described like the stationary case, and the short term fading is given by $\textbf{z}_{k,c} \sim \mathcal{CN}(\textbf{0},\,\textbf{I}_M)$. 
Thus, the channel from the UE $k$ is obtained:
\begin{equation}
\textbf{h}_k=\sum_{c=1}^{C_T} \textbf{h}_{k,c}
\end{equation}
where $C_T$ is the total number of clusters for the $k$th mobile user. Hence, the entire channel matrix  $\textbf{H}  \in \mathbb{C}^{M \times K}$ is defined simply as $\textbf{H} = [{\bf h}_1 \, {\bf h}_2\, \ldots  \,{\bf h}_k\, \ldots \,{\bf h}_K]$. 

\begin{algorithm}[!htbp]
\caption{Generation of the VR}\label{algo:VR}
\begin{algorithmic}[1]
 \STATE \textbf{Input} $M,\lambda,r_\text{min},r_\text{max}, p_0, p_1, c$
 \STATE Pick a radius $r$ with uniform probability distribution between $r_\text{min}$ and $r_\text{max}$
 \STATE $L_{\text{VR}} = 2 r$
 \STATE $d_H = 5 \lambda$
 \STATE $L_{\text{BS}} = (M-1) d_H$
 \STATE $M_{\text{VR}} =\left \lceil{\frac{M L_{\text{VR}}}{L_{\text{BS}}}}\right \rceil  $
 \STATE $\text{v}_{\text{prob}} = [p_0 \hspace{0.2 cm} p_1]$
 \STATE Pick a initial number $v_{\text{ant}}$ (0 or 1) with equal probability
 \STATE $i=0$
	\FOR{$n=1:M_{\text{VR}}$}
	\STATE $M_{\text{set}}(n)=v_{\text{ant}}$
 	\IF {$v_{\text{ant}} = 1$}
	\STATE $\text{v}_{\text{prob}} = [p_0 \hspace{0.2 cm} p_1]$
	\STATE $i=0$
	\ELSIF{$v_{\text{ant}} = 0$ and $(p_1 -i)\geq 0$}
	\STATE $\text{v}_{\text{prob}} = [p_1-i \hspace{0.2 cm} p_0+i]$
	\STATE $i=i+c$
	\ELSIF {$v_{\text{ant}} = 0$ and $(\text{pr}_1 -i)<0$}
	\STATE $\text{v}_{\text{prob}} = [0 \hspace{0.2 cm} 1]$
	\ENDIF
	\STATE Pick another number $v_{\text{ant}}$ (0 or 1) with probability function of $\text{v}_{\text{prob}}$
	\ENDFOR
\STATE $M_{\text{set}}$: antenna set within VR; antennas may be obstructed [$M_{\text{set}}(n)=0$] or not obstructed  [$M_{\text{set}}(n)=1$] 
\STATE \textbf{Output}  $M_{\text{set}}$
\end{algorithmic}
\end{algorithm}

\subsection{Downlink Transmission in XL-MIMO with Linear Precoding}
For the downlink transmission, the BS transmits payload data to its UEs, using a linear precoding. Let $s_k$ be the symbol to be transmitted to the $k$th user, where $\mathbb{E}\{|s_k|^2\}=1$. Each user is associated with the precoding vector $\textbf{w} \in \mathbb{C}^{M \times 1}$, that determines the spatial directivity of the transmission. Thus, the transmitted signal is described by:
\begin{equation}\label{eq:x}
\textbf{x}= \textbf{W}\textbf{s}
\end{equation}
\noindent where $\textbf{W} = [\textbf{w}_1 \hspace{0.2cm} ... \hspace{0.2cm} \textbf{w}_K]$ and $\textbf{s}$ is the transmitted signal vector. Aiming at satisfying the power constraints, the precoding must be chosen obeying $\mathbb{E}\{|\textbf{x}|^2\}=P$. The received signal at the $K$ users is given by:
\begin{equation}
\textbf{y}= \textbf{H}^H\textbf{x} + \textbf{n}
\end{equation}
\noindent where $(.)^H$ is the the Hermitian operator, $\textbf{n}$ is a vector whose $k$th element, $n_k$, is the additive noise at the $k$th user and $\textbf{H} \in \mathbb{C}^{M \times K}$ contains the channel vectors of each user. Two selected low-complexity linear precoding techniques are the Conjugate Beamforming (CB) and Zero-forcing (ZF), defined respectively by: 
\begin{equation}
\textbf{W}^{\text{CB}}=\textbf{H}
\label{eq:41}
\end{equation}
\begin{equation}
\textbf{W}^{\text{ZF}}=\textbf{H}(\textbf{H}^{H}\textbf{H})^{-1}
\label{eq:42}
\end{equation}

To satisfy the power constraint of the transmitted signal, the precoding in \eqref{eq:41} and \eqref{eq:42} are normalized as:
\begin{equation}
\textbf{W}^{\text{CB}}= \left[p_1 \frac{\textbf{w}_1^{\text{CB}}}{||\textbf{w}_1^{\text{CB}}||} \hspace{0.15cm} ... \hspace{0.15cm} p_K \frac{\textbf{w}_K^{\text{CB}}}{||\textbf{w}_K^{\text{CB}}||}\right] 
\label{eq:50}
\end{equation}

\begin{equation}
\textbf{W}^{\text{ZF}}=\left[p_1 \frac{\textbf{w}_1^{\text{ZF}}}{||\textbf{w}_1^{\text{ZF}}||}  \hspace{0.15cm} ...  \hspace{0.15cm} p_K \frac{\textbf{w}_K^{\text{ZF}}}{||\textbf{w}_K^{\text{ZF}}||}\right] 
\label{eq:50}
\end{equation}

\noindent where $p_1,...,p_K$ are the transmit powers for each user in which the average transmit power at the BS is $P=p_1+p_2+...+p_K$, $\textbf{w}_1^{\text{CB}},...,\textbf{w}_K^{\text{CB}}$ and $\textbf{w}_1^{\text{ZF}},...,\textbf{w}_K^{\text{ZF}}$ are the precoding vectors for each user according to the CB and ZF principle, respectively.

\section{Figures of Merit}\label{sec:FoM}
We analyze the channel models for M-MIMO and XL-MIMO presented previously using the capacity as a valid figure of merit. The ergodic channel capacity is given by:

\begin{equation}
C=\mathbb{E}\left\{ \log_2\left[\det\left(\textbf{I} + \frac{\eta}{M}\textbf{H}\textbf{H}^H\right)\right]\right\}
\label{eq:2}
\end{equation}
where $\eta$ is the average signal-to-noise ratio, $M$ is the number of antennas in BS and $\textbf{H}$ is the channel matrix. 

As the capacity is a concave function (logarithmic function), we can apply the Jensen's inequality to obtain the following upper bound (UB) on the mean (ergodic) capacity \cite{Loyka2001} .
\begin{equation}
C_{\text{ub}}=\log_2\left[\det\left(\textbf{I} + \frac{\eta}{M}\mathbb{E}\{ \textbf{R}\}\right)\right]
\label{eq:33}
\end{equation}

\noindent where $\textbf{R}$ is the channel matrix defined by each channel model. Thus, the upper bound on ergodic capacity $C_{\text{ub}}$ was analyzed to all channel models, assuming high signal-to-noise ratio.

To analyze the XL-MIMO channel models, we have selected the SINR as the figure of merit. The choice for another figure of merit to evaluate the XL-MIMO channel models was necessary due to the construction of the XL-MIMO model including the visibility regions, in step 2, and correlation in step 4,  Fig. \ref{fig:51}, where the correlation matrix is defined for a stationary case, and then added non-stationarity to the model. Thus, one can describe the SINR for the $k$th user as:
\begin{equation}
\gamma_k= \dfrac{ |\textbf{h}_k^{H} \textbf{g}_k|^2}{\sum\limits_{\substack{j=1\\ j\neq k}}^{K} |\textbf{h}_k^{H}\textbf{g}_j|^2 + \sigma^2}
\end{equation}

\noindent where $\sigma^2$ is the noise power, $\textbf{g}_k$ and $\textbf{h}_k$ are the precoding and channel vectors from user $k$, respectively

It is worth to note that across the numerical results section, the path loss of each XL-MIMO channel $\textbf{h}_{k,c}$ has been normalized by the inverse of path loss gain, i.e., $A = \widetilde{d}^\alpha$, defining the adopted power allocation policy. 

\section{Numerical Results}\label{sec:results}

In this section the validation of the analyzed M-MIMO and XL-MIMO channel models is corroborated numerically. The adopted channel and system parameter values  are described in Table \ref{tab:1}.

\begin{table}[!htbp]
\centering
\caption{Parameter values adopted in the numerical simulations for M-MIMO and XL-MIMO channels.}
\begin{tabular}{|l|c|}
\hline
\multicolumn{1}{|c|}{\textbf{Parameters}}                      & \textbf{Values}                \\ \hline
\multicolumn{2}{|c|}{\bf M-MIMO Channel}\\
	\hline
Average signal-to-noise ratio & $\eta = 60$ dB  \\
\hline
Figure of merit:  Ergodic capacity &  $C_{\rm ub}$\\
\hline
Path loss term                & $\beta$ = 1      \\ 
\hline
\multicolumn{2}{|c|}{\bf XL-MIMO Channel}\\
	\hline
	Average signal-to-noise ratio & $\eta = 10$ dB  \\ \hline
Figure of merit:  SINR & $\gamma_{k}$\\
	\hline
Number of antennas $M$ at the BS               & 100                \\ \hline
Number of clusters by UE                                      & 2                              \\ \hline
Distance $d_1$ for the 1st scheme of clusters distribution  & 35 m                           \\ \hline
Distance $d_2$ for the 2nd scheme of clusters distribution & 20 m                           \\ \hline
Distance between the UE and the BS $\widetilde{d}$ & 40 m                           \\ \hline
Path loss exponent for the VR                  & $\alpha_{\text{VR}}$= 3 dB                 \\ \hline
Path loss exponent out of the VR      & $\alpha_{\text{nVR}}$= 6 dB                 \\ \hline
Normalization factor   & $A = \widetilde{d}^{\alpha_{\text{VR}}} = 40^{3}$ \\ \hline
Cluster radius  $r\in [r_{\min};\,\, r_{\max}] $   & $r \sim \mathcal{U}(5,\,10)$ m \\ \hline
Probability of the antenna not visible                                & $p_0$ = 0.05 \\ \hline
Probability of the antenna is visible                                     & $p_1=1-p_0= 0.95$ \\ \hline
Factor c related to the VR                                & $c = 0.05$ \\ \hline
Channel average path loss gain $\mathcal{L}_0$              & $-34.53$ dB             \\ \hline
Pathloss reference distance & $d_0=1$ m\\ \hline
\multicolumn{2}{|c|}{\bf MIMO System}\\
	\hline
	Linear Precoding & CB;\,\, ZF \\ \hline
	Carrier frequency & $f=2.4$ [GHz] \\ \hline
Carrier wavelength & $\lambda = $12.5 [cm]\\ \hline
Total transmit power P               & 1 W                 \\ \hline
Transmit power for each user  $p_k$        & $P/K$               \\ \hline
Number of Monte Carlo realizations       & 300             \\ \hline
\end{tabular}
\label{tab:1}
\end{table}

\subsection{CBSM}\label{sec:cbsm}

{For CBSMs, the Table \ref{tab:2} describes the parameters evaluated in each simulation. Thus, we individually present the values used to generate each figure.}

\begin{table}[!htbp]
\centering
{\caption{Parameter values adopted for CBSMs.}
\label{tab:2}
\begin{tabular}{|l|c|c|c|c|c|c|}
\hline
\multicolumn{7}{|c|}{\textbf{CBSM - Parameters adopted in each simulation}}                                                             \\ \hline
   & Fig. 5a       & Fig. 5b   & Fig. 6a      & Fig. 6b    & Fig. 7a       & Fig. 7b     \\ \hline
Path loss term $\beta$                            & 1             & 1         & 1            & -          & 1             & 1           \\ \hline
Number of antennas $M$                            & {[}20:400{]}  & 100       & {[}20:400{]} & -          & {[}20:400{]}  & 100         \\ \hline
Correlation factor $\rho$                         & {[}0:0.2:1{]} & {[}0:1{]} & -            & -          & {[}0:0.2:1{]} & {[}0:1{]}   \\ \hline
Shadowing - standard deviation $\sigma_{\varphi}$ & -             & -         & {[}0:2:6{]}  & {[}0:10{]} & 4             & {[}0:2:6{]} \\ \hline
Angle of arrival $\theta$                         & -             & -         & -            & -          & $\pi / 2$     & $\pi / 2$   \\ \hline
\end{tabular}}
\end{table}

\subsubsection{Exponential}

\begin{figure}[!htbp]
\centering
\includegraphics[width=0.65\textwidth]{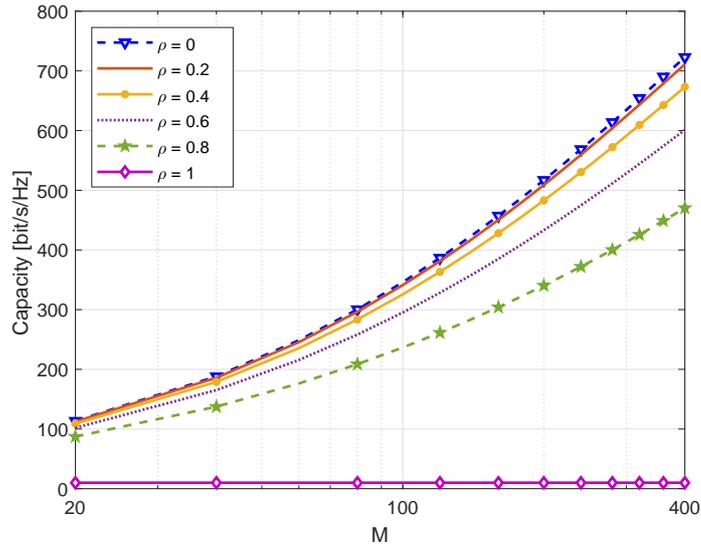}\\
{\bf a}) as a function of the number of antennas $M$;
\includegraphics[width=0.65\textwidth]{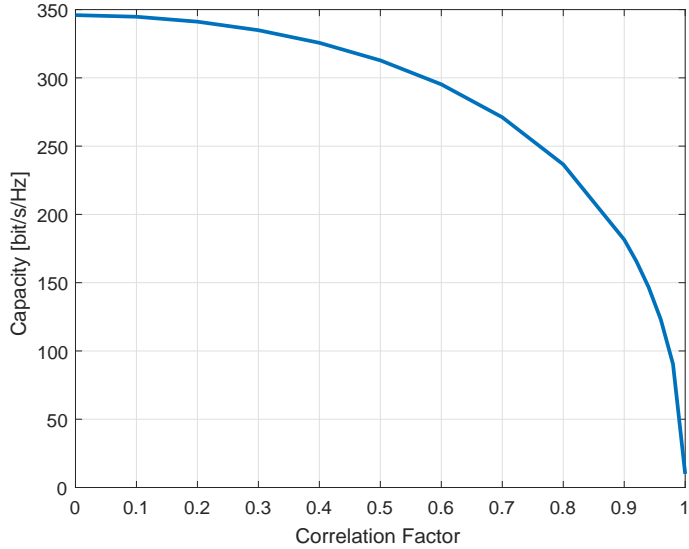}\\
{\bf b}) as a function of the correlation factor  ($M=100$).
\caption{CBSM exponential model. Capacity as a function of  a) number of antennas; (b) correlation factor assuming $M=100$.}
\label{fig:8}
\end{figure}

The first model is the exponential presented in eq. (\ref{eq:1}). The $C_{\text{ub}}$ for this model was analyzed by the number of antennas and by the increase of the correlation factor as showed in Fig. \ref{fig:8}.a and Fig. \ref{fig:8}.b, respectively.  The capacity presents an exponential gain with the logarithmic increase in the number of antennas. For this model, the number of BS antennas $M$ can increase the capacity up to 7 times, if we compare a system with 20 antennas against 400 antennas. However, the correlation factor has an undesirable effect on the capacity, mainly, when $\rho > 0.6$, the capacity decreases drastically. Therefore, it is possible to see that the correlation factor has a strong impact on the system capacity even under a large number of antennas.

\subsubsection{Uncorrelated Fading}

\begin{figure}[!htbp]
\centering
\includegraphics[width=0.49\textwidth]{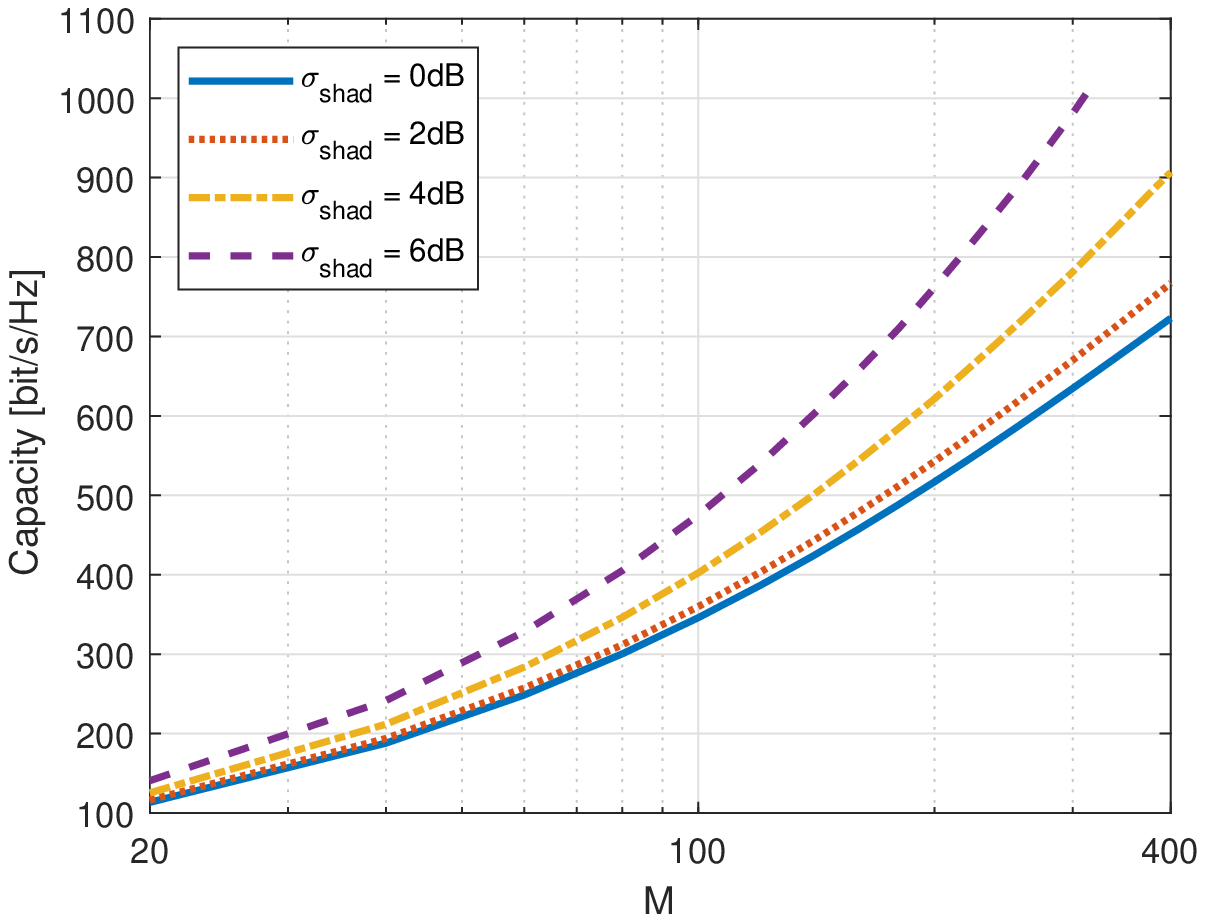}
\includegraphics[width=0.49\textwidth]{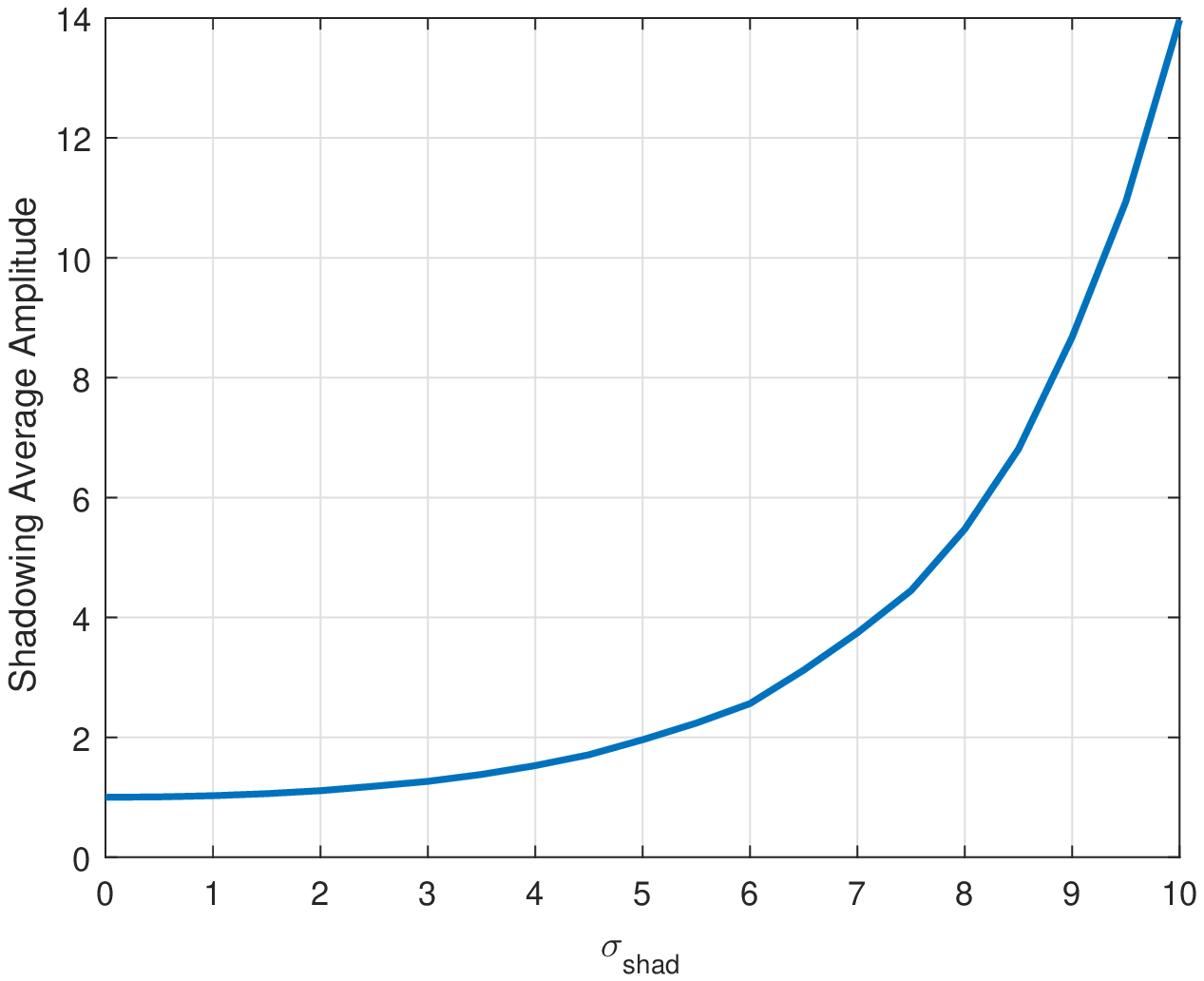}
{\bf a}) as a function of the number of antennas $M$; \hspace{20mm}{\bf b}) Shadowing analysis.\hspace{35mm}
\caption{Uncorrelated fading: a) Upper bound of capacity $\times$ number of antennas for different values of standard deviation of shadowing ($\sigma_{\text{shad}}$);  b) Shadowing amplitude when the standard deviation is increased.}
\label{fig:4}
\end{figure}

The second model is the uncorrelated fading defined in eq. (\ref{eq:4}). The analysis of capacity versus number of antennas is depicted in Fig. \ref{fig:4}.a. Besides the same increasing capacity behavior with the logarithmic increment on the number of antennas, one can observe the effect of the shadowing. Indeed, when the standard deviation of the shadowing increases, the channel capacity also increases.

In order to measure the gain that the shadowing can offer, in Fig. \ref{fig:4}.b, the shadowing average amplitude is depicted as a function of its standard deviation. One can see that the shadowing channel term is able to provide an exponential gain in the channel capacity explaining the behavior in the Fig. \ref{fig:4}. From the shadowing term described in eq. \eqref{eq:4}, one can verify that the term is always positive even when $f_m$ ($f_\text{1},...,f_\text{M}$) is negative. If $f_m$ has small variance, on the average, the shadowing term will be small. However, for high variances, $f_m$ may assume more likely larger values and, on average, the shadowing term will be large. Besides, the shadowing term provides a descorrelation effect between the antenna signals, increasing the channel capacity. The difference in capacity is high if we compare the channel with shadowing variance $\sigma_{\rm shad}$ of 1 dB against 10 dB, resulting in average amplitude of shadowing of 1 versus 14, respectively.

\subsubsection{Exponential Model with Large Scale Fading}
\begin{figure}[!htbp]
\centering
\includegraphics[trim={7mm 1mm 13mm 6mm},clip,width=0.65\linewidth]{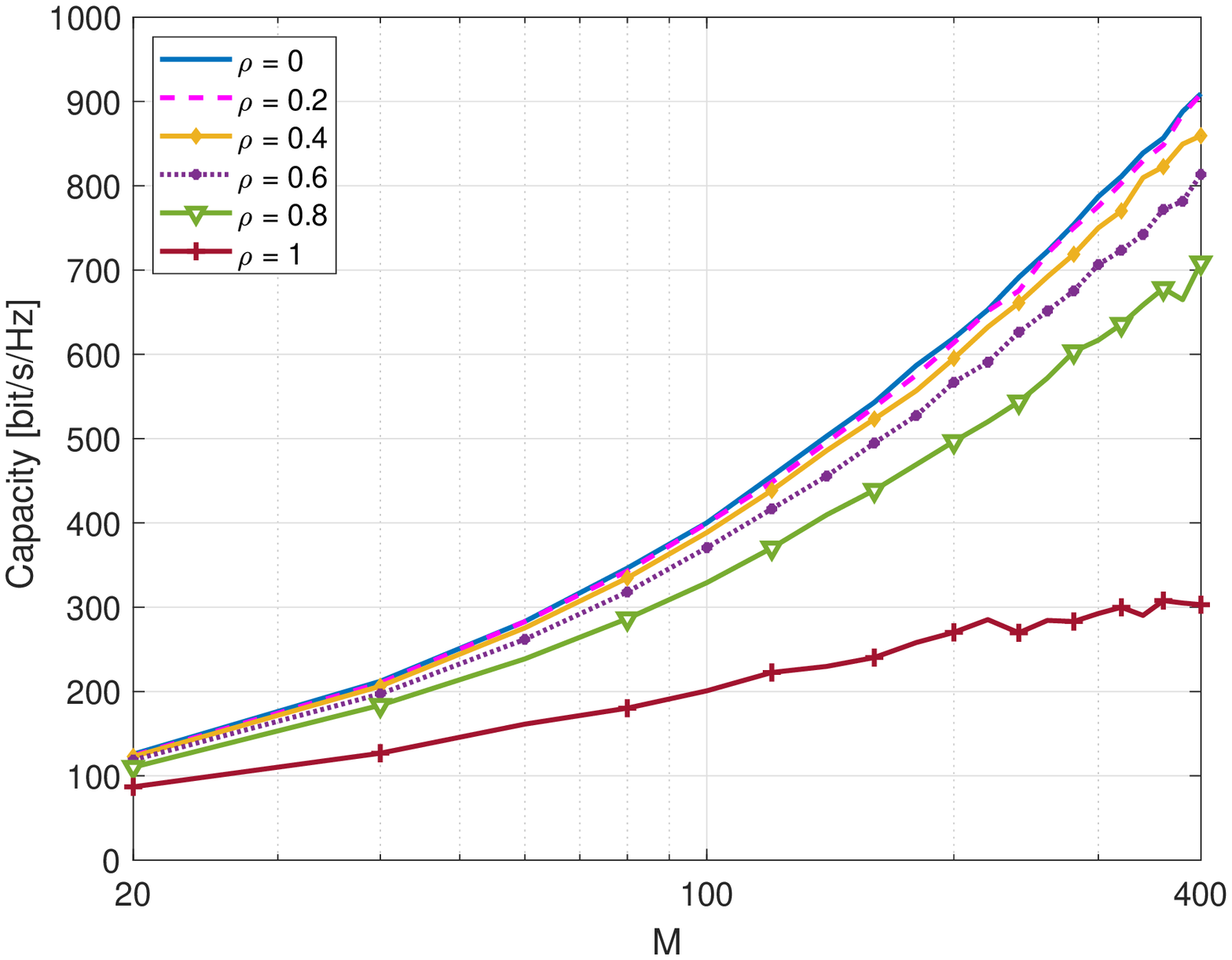}\\
{\bf a}) as a function of the number of antennas $M$ ($\sigma_{\rm shad}=4$dB)\\
\includegraphics[trim={7mm 1mm 13mm 6mm},clip,width=0.65\linewidth]{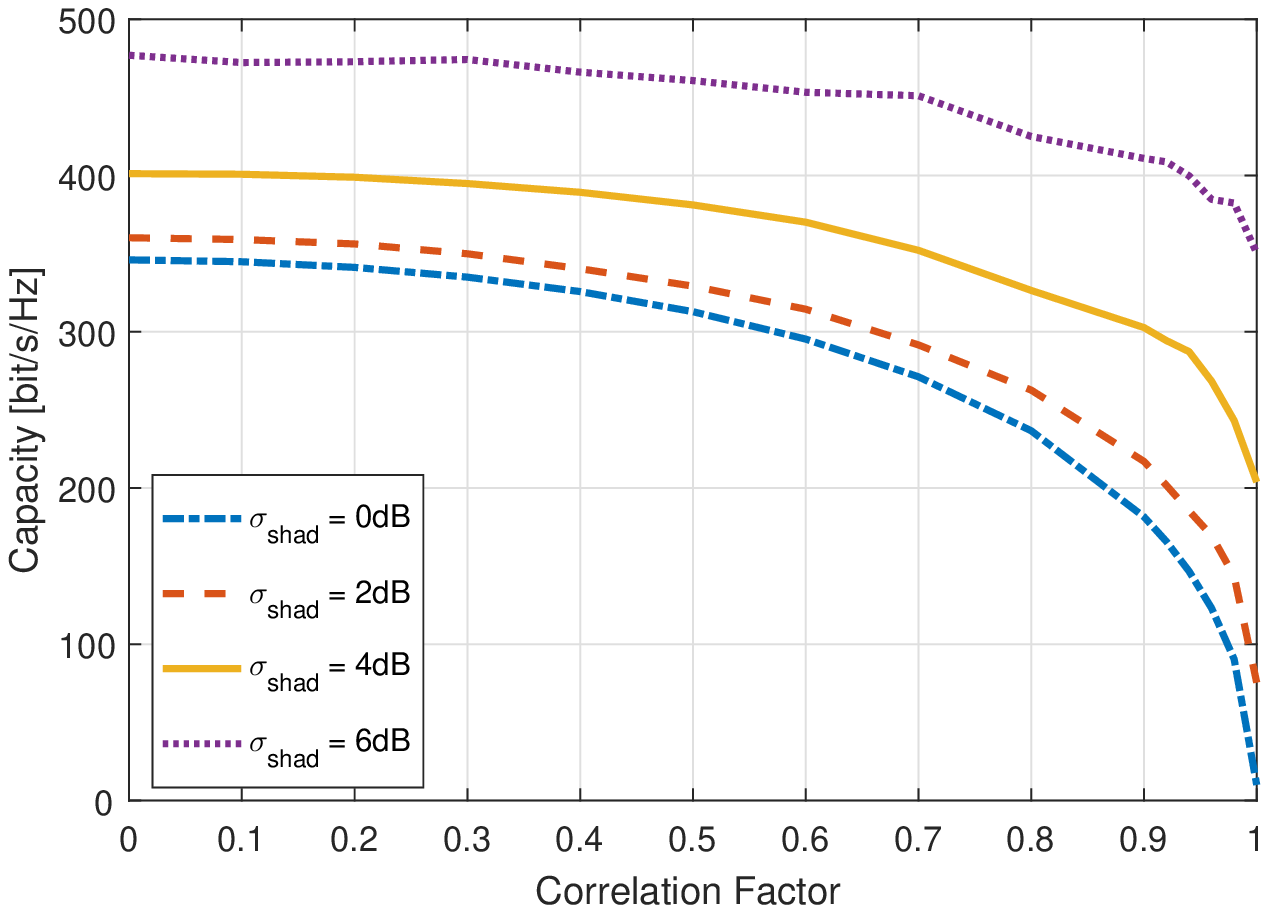}\\
{\bf b}) as a function of the correlation degree ($M=100$).
\caption{Correlated-based sthocastic} channel fading: a) UB capacity when the number of antennas increases; b) UB capacity for different values of std shadowing for $M=100$ antennas.
\label{fig:5a}
\end{figure}

The last CBSM considered herein is the exponential model combined with shadowing effects, as defined in eq. (\ref{eq:5}). The channel capacity for an increasing number of antennas and a growing correlation factor is depicted in Fig. \ref{fig:5a}.a and \ref{fig:5a}.b, respectively, where the increase rate of capacity with the number of antennas  depends on the  channel correlation index $\rho$.  However, differently from the simple exponential model, the values of capacity are higher due to presence of the shadowing. Moreover, one can observe that even under a correlation factor near or equal to $\rho=1$, the channel capacity increases steadily with $M$, but of course with reduced rate when $\rho \to1$. Furthermore, Fig. \ref{fig:5a}.b, shows channel capacity values for correlation values in the range $\rho \in [0; \,1]$ and  different standard deviation shadowing values. As expected, when the std shadowing values increases, the UB capacity increases too. Besides, notice that the capacity is close to uncorrelated channel capacity values without shadowing when the standard deviation value is 6 dB for a fully correlated channel condition, $\rho=1$.

\subsection{GBSM}
{For GBSMs, the Table \ref{tab:3} describes the parameter values adopted in each numerical simulation. Thus, we individually present the values used to generate each figure.}

\begin{table}[!htbp]
\centering
{\caption{Adopted values for GBSMs channel parameters.}
\label{tab:3}
\small
\begin{tabular}{|c|c|c|c|c|c|c|c|c|c|}
\hline                                                                                                                                                                                                                                                                                                                                                                                                           
\multicolumn{1}{|l|}{\rotatebox{90}{\bf Figure}} & \multicolumn{1}{l|}{\rotatebox{90}{\# antennas $M$}} & \multicolumn{1}{l|}{\rotatebox{90}{Antenna spacing $d_H$}} & \multicolumn{1}{l|}{\rotatebox{90}{AoA (azimuth) $\varphi$}} & \multicolumn{1}{l|}{\rotatebox{90}{AoA Range $\Delta_{\varphi}$}} & \multicolumn{1}{l|}{\rotatebox{90}{AoA Std $\sigma_{\varphi}$}} & \multicolumn{1}{l|}{\rotatebox{90}{Shad. Std $\sigma_{\text{shad}}$}} & \multicolumn{1}{l|}{\rotatebox{90}{AoA (elevation) $\theta$}} & \multicolumn{1}{l|}{\rotatebox{90}{AoA Range $\Delta_{\theta}$}} & \multicolumn{1}{l|}{\rotatebox{90}{AoA Std $\sigma_{\theta}$}} \\ \hline\hline
 8b                & 100                                           & 0.5 $\lambda$                                & {[}0;360)                                                   & {[}0:45{]}                                                         & -                                                        & -                                                             & -                                                            & -                                                                 & -                                                       \\ \hline
 9a                & 100                                           & 0.5 $\lambda$                                & $\pi/6$                                                     & {[}1:50{]}                                                         & -                                                        & -                                                             & -                                                            & -                                                                 & -                                                       \\ \hline
 9b                & 100                                           & 0.5 $\lambda$                                & $\pi/6$                                                     & $\small \sqrt{3}[10; 20; 30]$                                              & -                                                        & -                                                             & -                                                            & -                                                                 & -                                                       \\ \hline
 10a               & 100                                           & 0.5 $\lambda$                                & {[}0;90{]}                                                  & {[}0:45{]}                                                         & -                                                        & -                                                             & -                                                            & -                                                                 & -                                                       \\ \hline
 10b               & 100                                           & 0.5 $\lambda$                                & {[}0:360{]}                                                 & {[}10;30{]}                                                        & -                                                        & -                                                             & -                                                            & -                                                                 & -                                                       \\ \hline
 10c               & [20:400]                                        & 0.5 $\lambda$                                & {[}0;90{]}                                                  & {[}10;30{]}                                                        & -                                                        & -                                                             & -                                                            & -                                                                 & -                                                       \\ \hline
 10d               & 100                                           & {[}0:10{]} $\lambda$                         & 0                                                           & {[}10;30{]}                                                        & -                                                        & -                                                             & -                                                            & -                                                                 & -                                                       \\ \hline
 11a               & 100                                           & 0.5 $\lambda$                                & $\pi/6$                                                     & -                                                                  & {[}0:5:15{]}                                             & 0                                                             & -                                                            & -                                                                 & -                                                       \\ \hline
 11b               & 100                                           & 0.5 $\lambda$                                & $\pi/6$                                                     & -                                                                  & {[}0:5:15{]}                                             & 2                                                             & -                                                            & -                                                                 & -                                                       \\ \hline
 12a               & 100                                           & 0.5 $\lambda$                                & {[}0;90{]}                                                  & -                                                                  & {[}0:15{]}                                               & {[}0:2:4{]}                                                   & -                                                            & -                                                                 & -                                                       \\ \hline
 12b               & 100                                           & 0.5 $\lambda$                                & {[}0:360{]}                                                 & -                                                                  & 15                                                       & {[}0:2:4{]}                                                   & -                                                            & -                                                                 & -                                                       \\ \hline
 12c               & {[}20:400{]}                                  & 0.5 $\lambda$                                & {[}0;90{]}                                                  & -                                                                  & 15                                                       & {[}0:2:4{]}                                                   & -                                                            & -                                                                 & -                                                       \\ \hline
 12d               & 100                                           & {[}0:10{]} $\lambda$                         & 0                                                           & -                                                                  & {[}5;15{]}                                               & 0                                                             & -                                                            & -                                                                 & -                                                       \\ \hline
 13a               & 100                                           & 0.5 $\lambda$                                & {[}0:360{]}                                                 & 10                                                                 & -                                                        & -                                                             & {[}-90:90{]}                                                 & 2                                                                 & -                                                       \\ \hline
 13b               & {[}16:400{]}                                  & 0.5 $\lambda$                                & {[}0;90{]}                                                  & 30                                                                 & -                                                        & -                                                             & {[}0;90{]}                                                   & {[}15;30{]}                                                       & -                                                       \\ \hline
 13c               & 100                                           & 0.5 $\lambda$                                & {[}0:360{]}                                                 & {[}10;30{]}                                                        & -                                                        & -                                                             & 0                                                            & {[}2;10;30{]}                                                     & -                                                       \\ \hline
 13d               & 100                                           & {[}0:10{]} $\lambda$                         & 0                                                           & {[}10;40{]}                                                        & -                                                        & -                                                             & 0                                                            & {[}5;20{]}                                                        & -                                                       \\ \hline
 14a               & {[}16:400{]}                                  & 0.5 $\lambda$                                & {[}0;90{]}                                                  & -                                                                  & 30                                                       & -                                                             & {[}0;90{]}                                                   & -                                                                 & {[}15;30{]}                                             \\ \hline
 14b               & 100                                           & 0.5 $\lambda$                                & {[}0:360{]}                                                 & -                                                                  & {[}10;30{]}                                              & -                                                             & 0                                                            & -                                                                 & {[}2;10{]}                                              \\ \hline
 14c               & 100                                           & {[}0:10{]} $\lambda$                         & 0                                                           & -                                                                  & {[}5;20{]}                                               & -                                                             & 0                                                            & -                                                                 & {[}5;20{]}                                              \\ \hline
\end{tabular}}
\end{table}

\subsubsection{One-ring Model with ULA}
\begin{figure}[!htbp]
\centering
\includegraphics[trim={1mm 1mm 2mm 2mm},clip,width=0.44\linewidth]{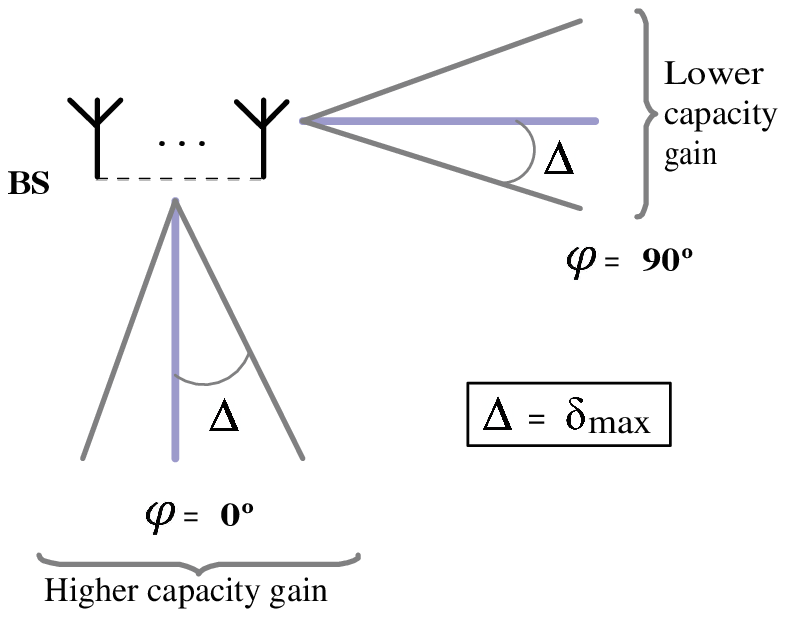} \quad
\includegraphics[trim={5mm 6mm 7mm 8mm},clip,width=0.52\linewidth]{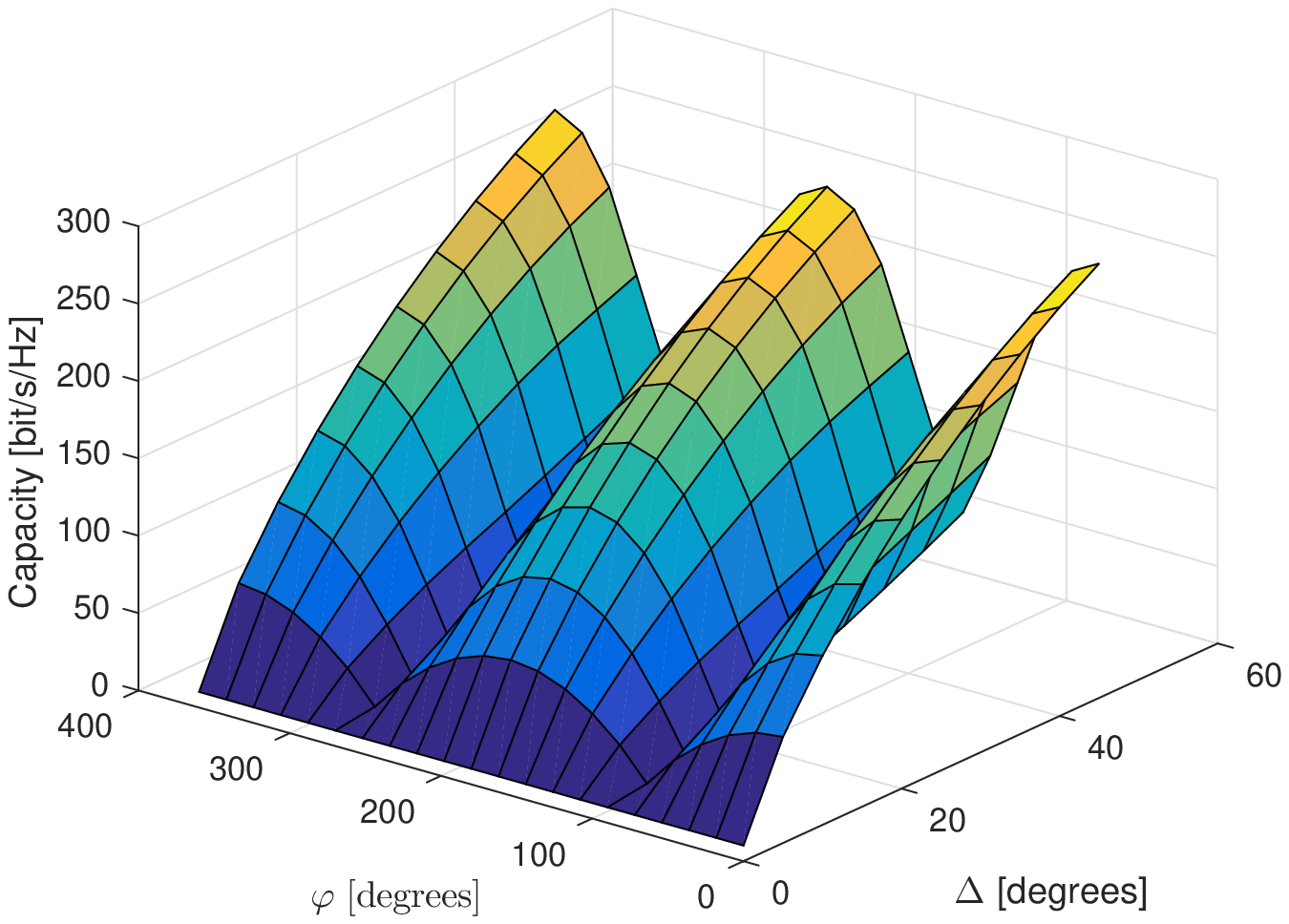}\\
{\bf a}) Relation between capacity and nominal AoA; \quad \quad {\bf b}) Capacity  {\it vs}  AoA $\varphi$ and angular spread $\Delta$ ($M=100$, $d_H=0.5$)
\caption{Capacity analysis in geometric-based stochastic models (GBSM)  according to the AoA considering ULA arrangements. a) Illustration of the relation between capacity and nominal AoA.  b) Capacity analysis  {\it vs} AoA and $\Delta$.}
\label{fig:31}
\end{figure}

Next, we analyze the channel models for M-MIMO and XL-MIMO systems based on the geometry, starting with the 2D models for scenarios where the user is around the scatterers uniformly distributed \textit{i.e.} the one-ring geometry, as defined in eq. (\ref{eq:12}).  First, the capacity was analyzed when the nominal AoA $\varphi$ and the angular spread $\Delta$ are varying, as illustrated in Fig. \ref{fig:31}.  Indeed, if we imagine a horizontal ULA, we can suppose that perpendicular angle of arrival will present higher capacity than parallel ones, because parallel angles result in antennas strongly correlated, hence, smaller capacity.

This behaviour is observed in Fig. \ref{fig:31}.b, in which the angles $0^o$ and $180^o$ represent the perpendicular angles to the antennas array and the angles $90^o$ and $270^o$ represent the parallel angles to the antennas. We also observed that the angular spread $\Delta$ can result in an increase in the capacity. The $\Delta$ is related to how close the angle of the signals received from each multipath component are, indicating that the arrived signals may are correlationed according to the value of $\Delta$. 

\begin{figure}[!htbp]
\centering
\includegraphics[width=0.49\textwidth]{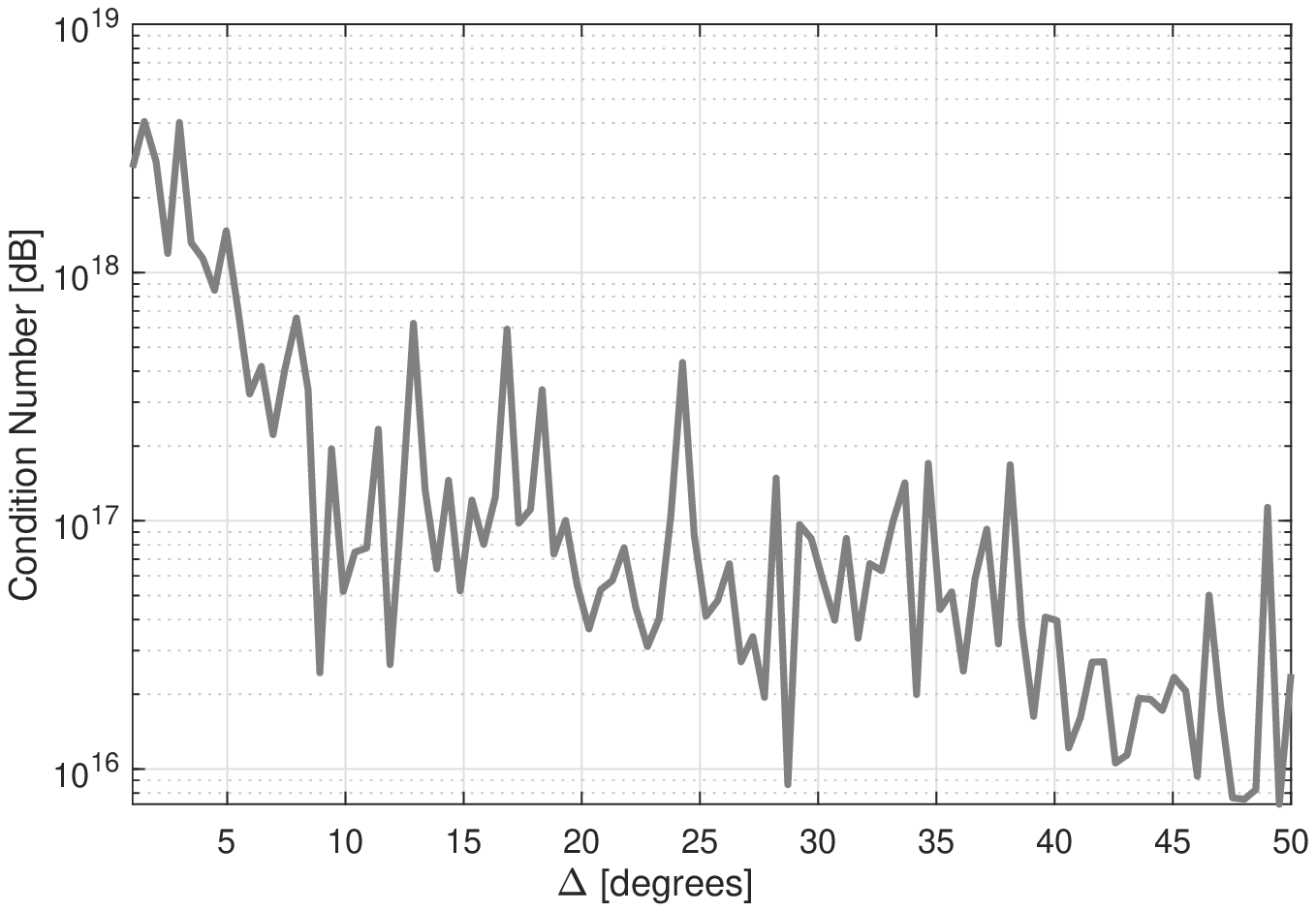}
\includegraphics[width=0.49\textwidth]{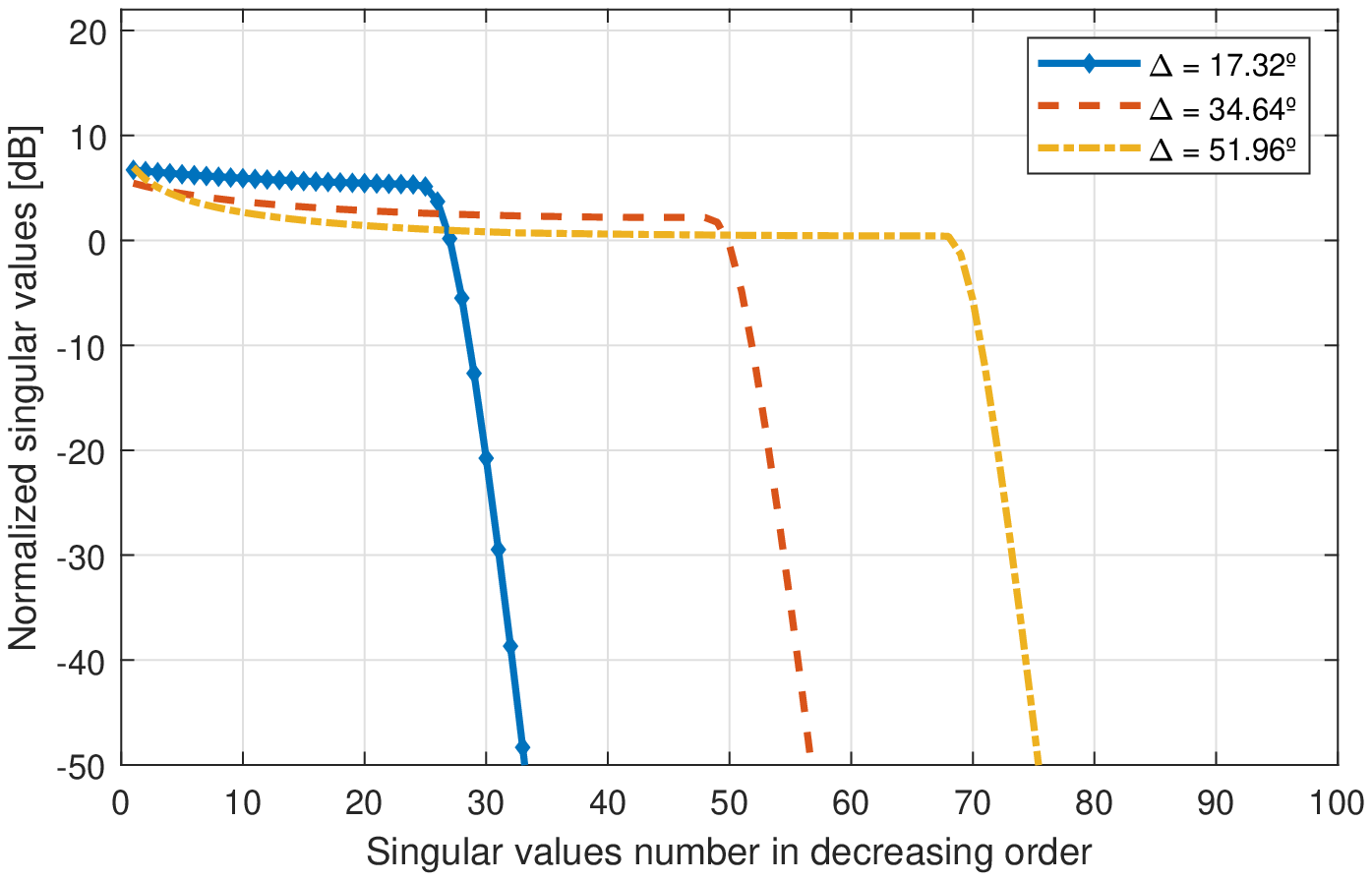}
{\bf a}) as a function of $\Delta$; \hspace{55mm}{\bf b}) Singular values decomposition analysis.\hspace{10mm}
\caption{a) Condition number when the angular spread is increased; b) Singular values for different angular spread.}
\label{fig:20} 
\end{figure}

Aiming to verify if the behavior of capacity according to $\Delta$ is due to the spatial correlation, we analyzed the \textit{condition number} (CN), {\it i.e.} the ratio between the maximum ($\sigma_{\max}$)  and the minimum $(\sigma_{\min}$) singular value related to the channel matrix $\bf A$ \cite{Belsley1980}:
\begin{equation}
\kappa(\textbf{A})=\frac{\sigma_{\max}}{\sigma_{\min}}
\label{eq:37}
\end{equation}
In our massive MIMO channel analysis, the CN is used as an environmental characterization metric and it can express the spatial correlation degree of a correlation channel matrix \cite{costa2011}. As well known, if a matrix is correlationed, only the first singular value is representative, while the others result in ''small'' singular values, indicating that the matrix is near to the singularity.  Hence, in Fig. \ref{fig:20}, one can see that the CN has decreased with the increasing of $\Delta$; however, such values remain high due to the high correlation associated with the One-ring channel model. Therefore, the behavior seen in Fig. \ref{fig:31}.b with respect to $\Delta$ is due to the degree of correlation in the One-ring model, in which the higher the degree of correlation, the lower capacity.

\begin{figure}[!htbp]
\centering
\includegraphics[width=0.49\textwidth]{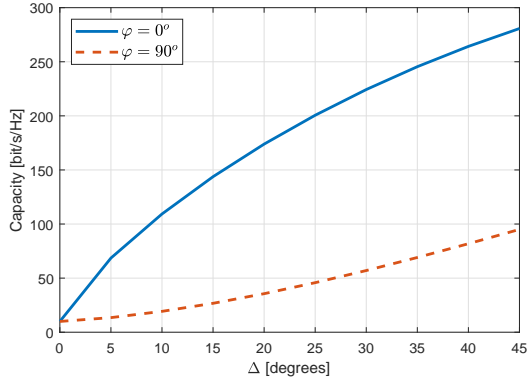}
\includegraphics[width=0.49\textwidth]{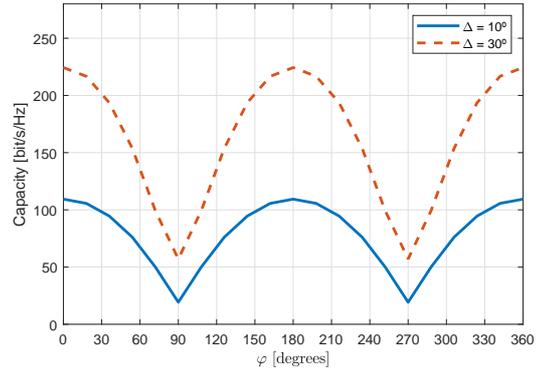}\\
{\bf a}) as a function of the $\Delta$ \hspace{60mm} {\bf b}) as a function of nominal AoA \\
\includegraphics[width=0.49\textwidth]{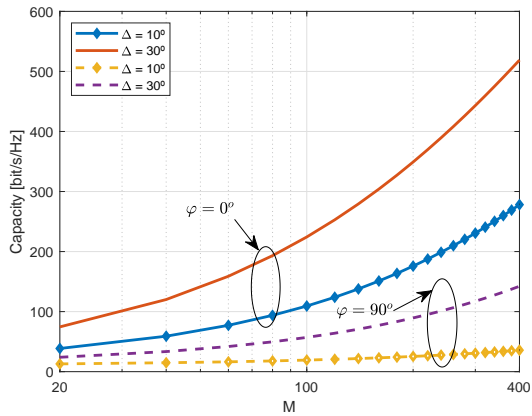}
\includegraphics[width=0.49\textwidth]{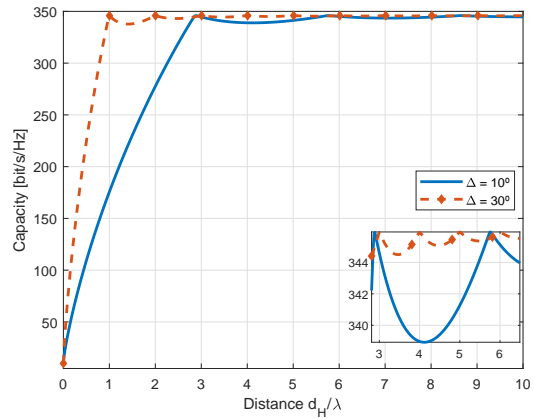}\\
{\bf c})  increasing number of antennas ($d_H=0.5$) \hspace{20mm}  {\bf d}) normalized distance between the antennas ($\varphi = 0^o$) 
\caption{One-ring model with $M=100$ antennas. Capacity analysis  according to a) $\Delta$; b) AoA;  c) number of antennas; d) distance between antennas.}
\label{fig:14}
\end{figure}

In order to compare the capacity values of the GBSM and CBSM models, Fig. \ref{fig:14} depicts  the variation of capacity with $\Delta$, AoA, number of antennas $M$ and normalized distance between antennas $\frac{d_H}{\lambda}$. Fig. \ref{fig:14}.a analyzes the capacity as a function of $\Delta$ assuming fixed the nominal angles of arrival in $\varphi=0^o$ and $90^o$, representing the best and worst case in terms of capacity for a wide range of $\Delta$ values. Hence,  from the two AoA values, one can see that to $\Delta$ near zero, the capacity is small and approximately constant, \textit{ i.e.} independent of the nominal angle of arrival $\varphi$. However, one can see a huge difference of almost 190 bit/s/Hz with $\Delta = 45^o$. Moreover, even with the increase of $\Delta$, nominal angles of $90^o$ and $270^o$ do not increase significantly the capacity when compares to $0^o$ and $180^o$. Such behavior is justified if we imagine two arrival angles $\varphi = 0^o$ and $\varphi = 90^o$; in the first case, the ULA would receive the signals with the best AoA and the in the second case the ULA would receive the signals with the worst AoA, thus, even as the $\Delta$ increases, the capacity will increase much fast for the first case than for the second case. Moreover, comparing GBSM with CBSM model (section \ref{sec:cbsm}), one can observe a similarity in GBSM behavior with respect to $\Delta$ and the correlation factor presented in \ref{fig:8}.b), where the correlation factor and $\Delta$ are inversely proportional. Capacity behavior according to the number of antennas was analyzed in Fig. \ref{fig:14}.c). For any angle of arrival and with a nonzero $\Delta$, the capacity increases with the number of antennas. However, the observed capacity values are very low if compared with the Exponential model of Fig. \ref{fig:8}.a). According to the geometric model, the obtained capacity does not exceed 500 bit/s/Hz, being this value compared to the CBSM with a correlation factor between $0.8$ and $1$. Moreover, we see that the difference in the capacity is more pronounced between the angles $\varphi$ when the number of antennas increases. The last analysis was developed according to the distance between the antennas presented in Fig.  \ref{fig:14} d). As the distance increased, we observed that capacity increased, and capacity remained constant over a certain distance. This behavior is related to the degree of correlation between the antennas, where with increasing spacing, the degree of correlation decreases. The fact that capacity reaches a limit indicates that the antennas are completely uncorrelated. The difference between $\Delta=10^o$ and $\Delta=30^o$ is that under the latter angular spread the signals are more uncorrelated, thus, the capacity increases with a distance smaller than $\Delta=10^o$.

\subsubsection{Gaussian Model with ULA}
\begin{figure}[!htbp]
\centering
\includegraphics[width=0.47\textwidth]{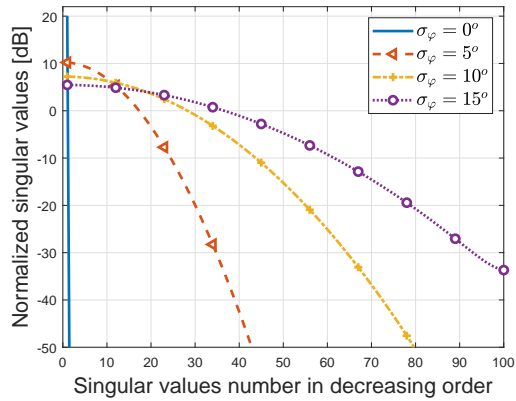}
\includegraphics[width=0.47\textwidth]{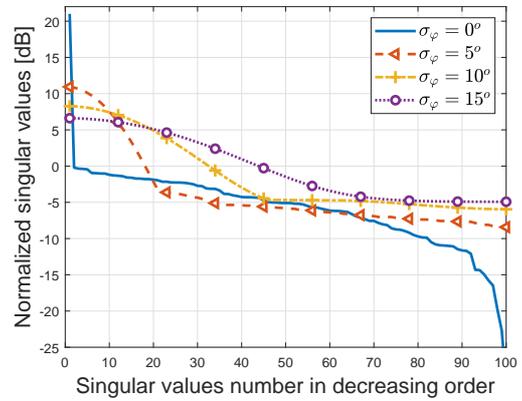} \\
{\bf a}) Singular values for shadowing  $\sigma_{\rm shad} = 0$ dB; \hspace{15mm}{\bf b}) Singular values for shadowing $\sigma_{\rm shad} = 2$ dB.
\caption{Singular values decomposition analysis for the  GBSM Gaussian local scattering model}.
\label{fig:24}
\end{figure}

The second GBSM is the Gaussian Local Scattering Model defined in eq. (\ref{eq:15}). Unlike the One-ring model, in the Gaussian model the spreaders are around to the user with Gaussian distribution, suggesting that in a propagation environment some spreaders are more likely to influence signals arriving at the BS. The first analysis was performed by analyzing the capacity when the $\sigma_{\varphi}$ increases. Such behavior is similar to $\Delta$ and it was explained for the One-ring model using CN metrics. Even knowing that $\sigma_{\varphi}$ affects the degree of correlation between signals, we analyze the singular values, but in order to verify the influence of shadowing on signal correlation. In Fig. \ref{fig:24}, there are two cases, the first without the shadowing effect, Fig. \ref{fig:24}.a, and the second with shadowing of 2 dB, Fig. \ref{fig:24}.b. Clearly one can see that increasing shadowing standard deviation implies in decreasing the difference between the largest singular value and the smallest singular value, resulting in a better (smaller) condition number. Such result is understandable if we realize that each antenna receives a signal with a shadowing amplitude distinct from the shadowing amplitude of another antenna, such amplitude being given by a random variable and uncorrelated to the other one\footnote{See the definition of correlation elements in the Gaussian Local Scattering Model, eq. (\ref{eq:15}), as well as the random fluctuation of the large-scale fading $f_m$ description in eq. \eqref{eq:4}}.

\begin{figure}[!htbp]
\centering
\includegraphics[width=0.495\textwidth]{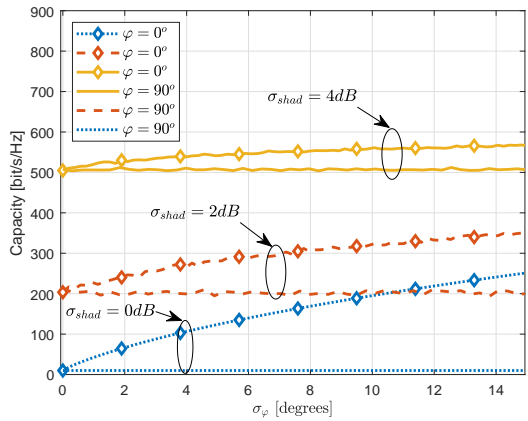} 
\includegraphics[width=0.49\textwidth]{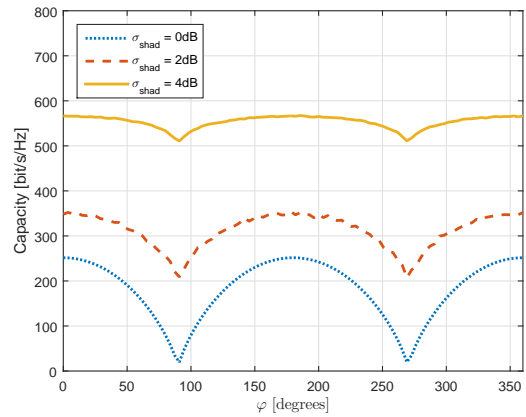}
{\bf a}) ASD with  $M=100$. \hspace{55mm}  {\bf b}) AoA, with $M=100$ and $\sigma_{\varphi}=15^{o}$ \\
\includegraphics[width=0.495\textwidth]{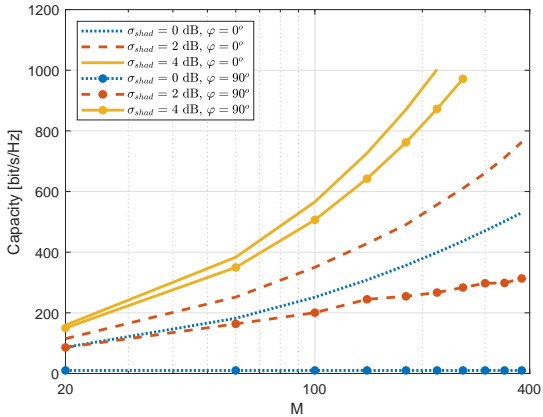}
\includegraphics[width=0.495\textwidth]{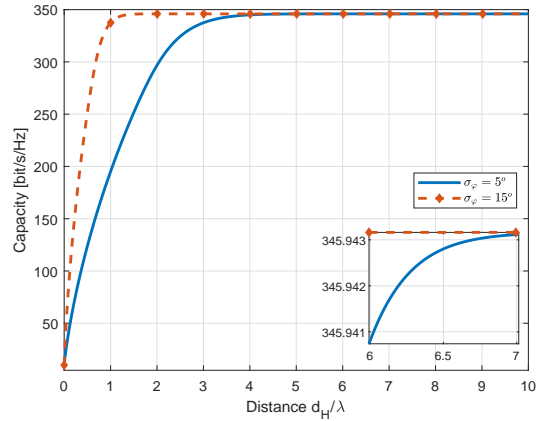}\\
{\bf c}) increasing number of antennas with $\sigma_{\varphi}= 15^{o}$. \hspace{20mm}  {\bf d}) Distance between antennas \hspace{25mm} 
\caption{Gaussian Local Scattering Model (approximate expression). Capacity analysis  according to: a) ASD;   b) AoA; c) increasing $M$; d) normalized distance between antennas.}
\label{fig:17}   
\end{figure}

In Fig. \ref{fig:17} one can check the capacity value for $M=100$ antennas under the Gaussian local scattering model with ULA when varying the angular standard deviation (ASD, $\sigma_{\varphi}$) and AoA. One can infer the gain that ASD provides; however, at angles close to $\varphi=90^o$ and $\varphi=270^o$ the models suggest that there is no gain (eq. (\ref{eq:15})). Thus, the capacity remains constant regardless of the presence of shadowing. We also see that shadowing has a greater influence on capacity than the increase in ASD itself; so, we can see that for ASD equal to zero we have a capacity gain of 500 bit/s/Hz with a shadowing standard deviation of 4 dB, \textit{i.e.}, even though both variables result in a reduction in the degree of correlation, shadowing has a greater impact than angular scattering. Also, we have observed that due to the signals from each antenna are more uncorrelated when the shadowing value is higher, the impact of the arrival angle on the degree of signal correlation is smaller, resulting in less impact on capacity. 

Another analysis for the Gaussian model is related to the increase of the number of antennas presented in Fig. \ref{fig:17}.c. For this case, we also check the dependence with three shadowing values. The capacity gain was exponential for this model as well as the previous models. If we compare the capacity gain provided by this model without shadowing against the One-ring model, one can see that both models are similar, despite the Gaussian model resulting in a greater gain that presented by the One-ring. channel model. Finally, Fig. \ref{fig:17}.d demonstrates that the channel capacity under the Gaussian model in ULA antenna arrangement increases when distance between antenna array elements increases as well, but unlike the One-ring model, at a certain distance the capacity remains constant without slight oscillations. Anyway,  in both channel models when the capacity is stabilized with antenna distance increasing, such capacity values become very close. 

\subsection{3D Models}

\subsubsection{One-ring Model with UPA}
\begin{figure}[!htbp]
\centering
\includegraphics[width=0.495\textwidth]{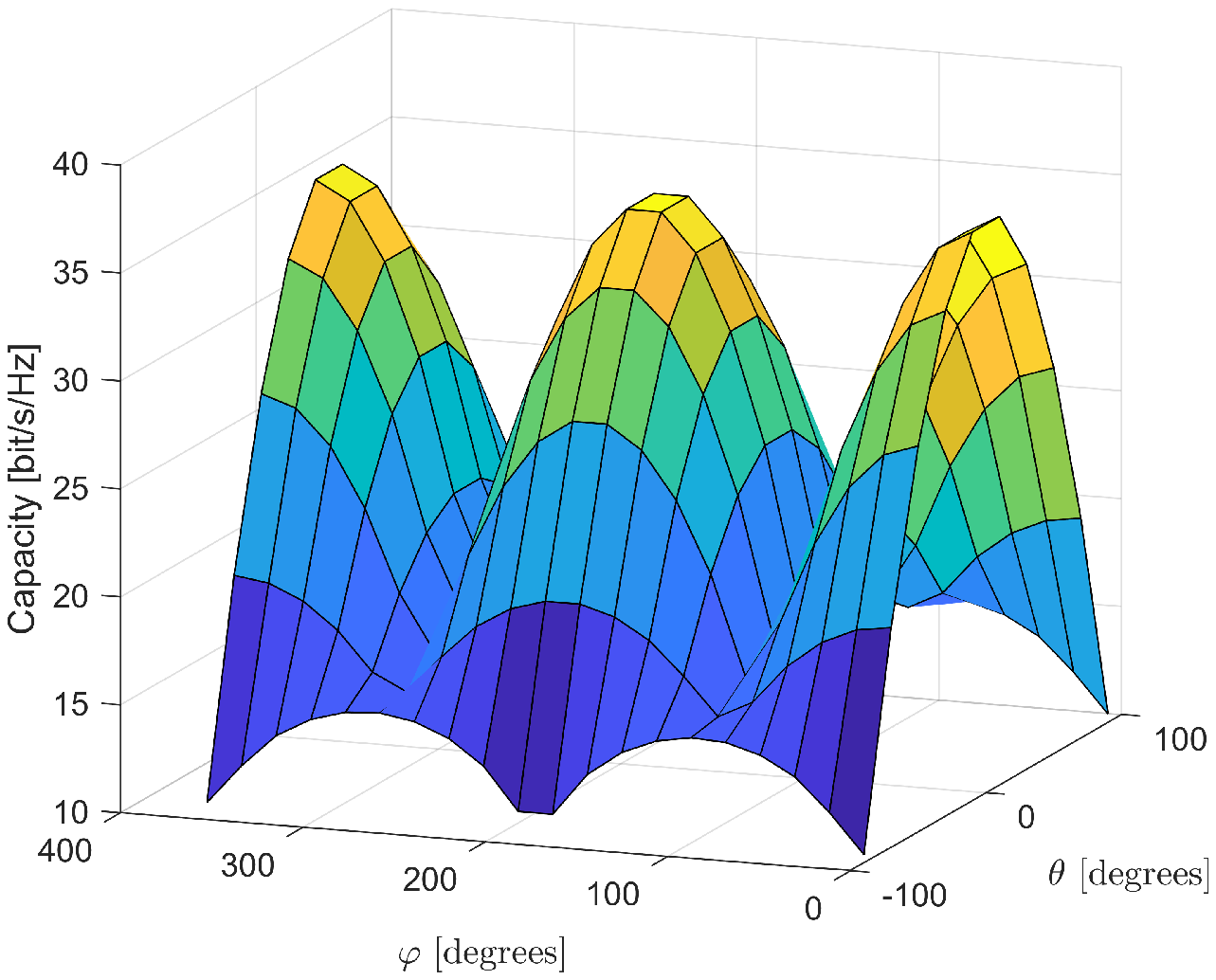}
\includegraphics[width=0.495\textwidth]{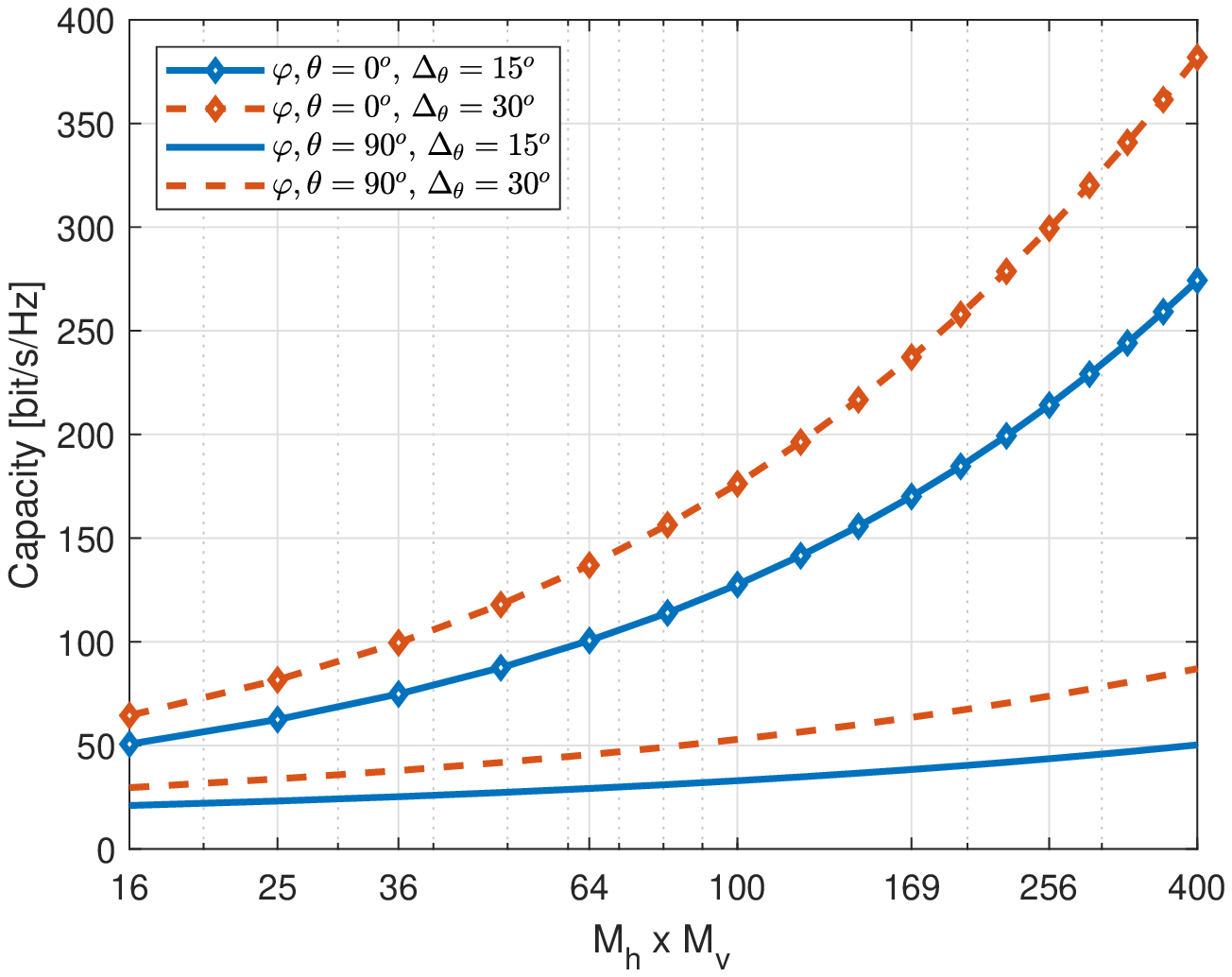}\\
{\bf a}) nominal AoA ($\Delta_{\varphi} = 10^o$ and $\Delta_{\theta} = 2^o$) \hspace{15mm}  {\bf b}) number of antennas  ($\Delta_\varphi = 30^o$). \\
\includegraphics[width=0.495\textwidth]{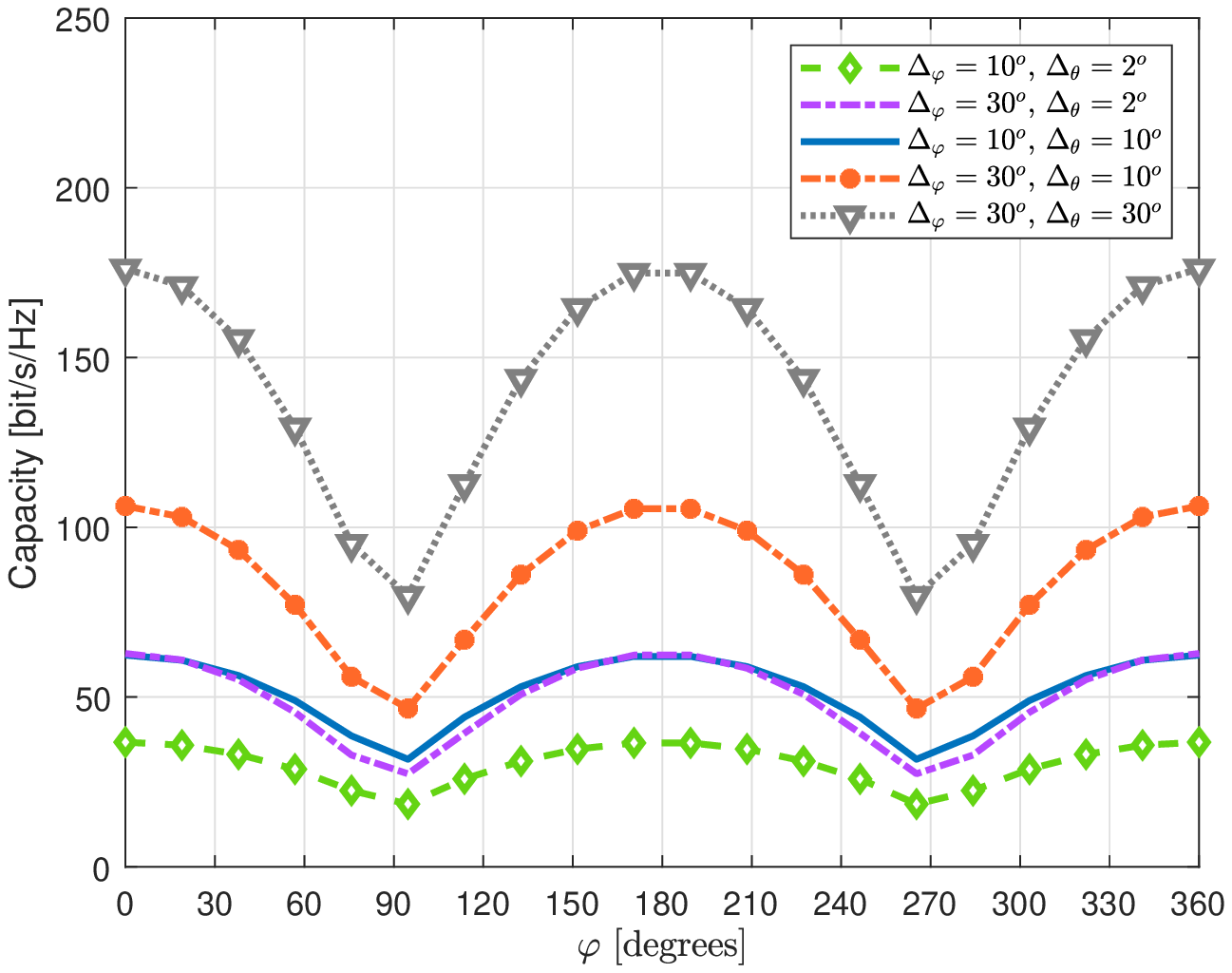}
\includegraphics[width=0.495\textwidth]{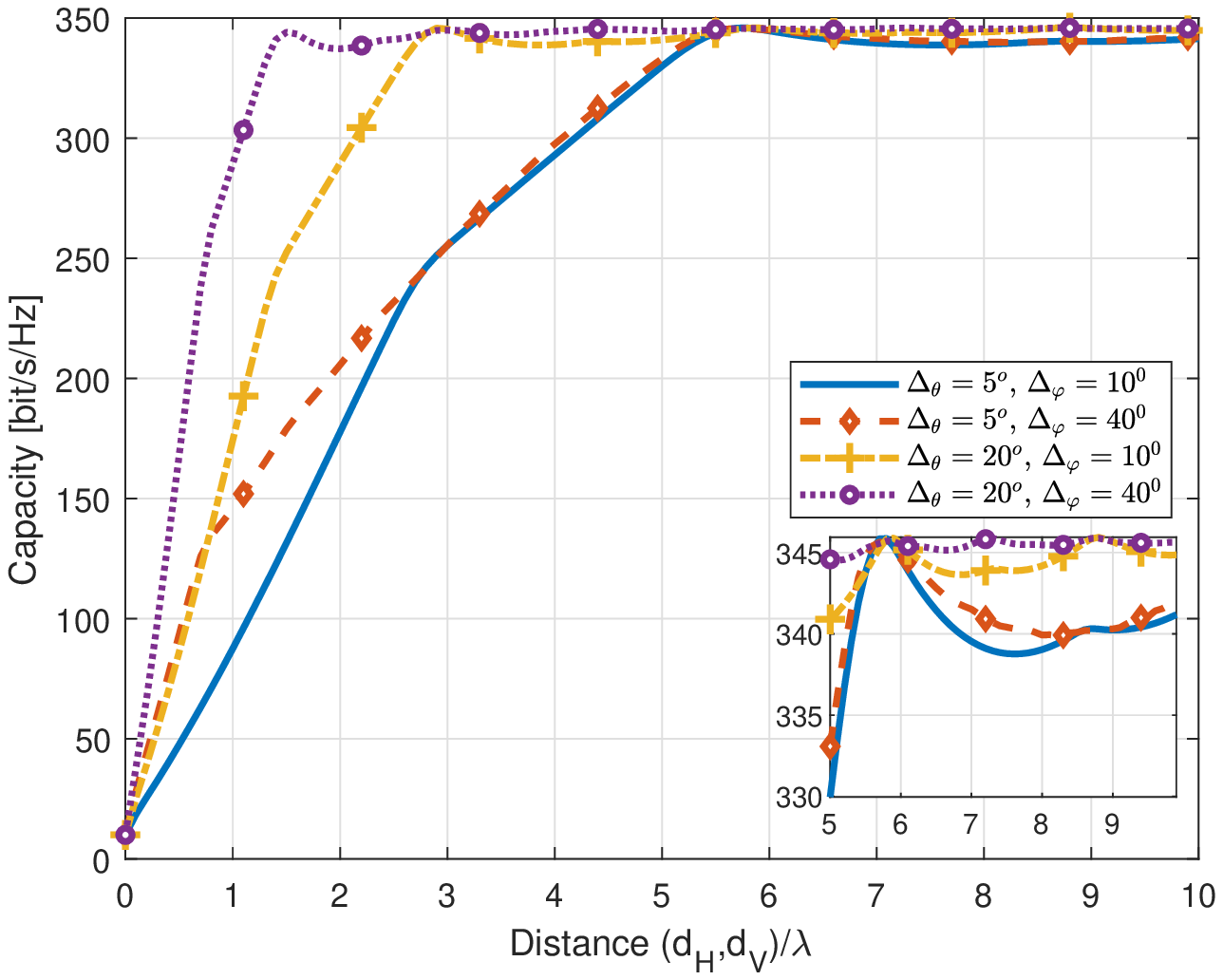}\\
{\bf c}) azimuth nominal AoA ($M=100$, $\theta = 0^o$) \hspace{15mm}  {\bf d}) as a function of the spacing between the antennas ($\varphi,\theta=0^o$) \\
\caption{One-ring Model with UPA. Capacity analysis according to: a) the nominal angles of azimuth and elevation; b) increasing $M_\text{h}$ and $M_\text{v}$ simultaneously; c) the azimuth nominal angle; d) increasing the horizontal and vertical spacing between the antennas.}
\label{fig:22}
\end{figure}

The next analysis is related to GBSM with UPA. In a model where the antennas array configuration is planar, we have to analyze not only the azimuth angle, $\varphi$, but also the elevation angle, $\theta$. First, the capacity behavior is presented according to the nominal angles in Fig. \ref{fig:22}.a. For the azimuth angles, the better arrival angles are the same demonstrated to ULA. Indeed, using ULA, one can identify that the elevation angles is not a determinant factor on the system performance, while in UPA the elevation angle affects greatly the channel capacity. As one can suppose or infer, the better angle is perpendicular to the azimuth and the elevation \textit{i.e.} when both is equal to  $0^{\rm o}$, the capacity is maximum, as depicted in Fig. \ref{fig:22}.a. Capacity behavior according to the number of antennas is also analyzed and presented in Fig. \ref{fig:22}.b, when the nominal angles are equal to $0^o$ (best case) and equal to $90^o$ (worst case). Previously, we saw that the UPA is limited by elevation and azimuth angles while ULA is only limited by azimuth angle. In fact, the UPA presents the advantage to perform the beamforming and a more versatile and compact spatial arrangement of antennas, but the capacity is smaller than ULA; for instance, one can see that the capacity  value is $C \approx 100$ bit/s/Hz when the BS is equipped with an UPA arranged as $M_{\rm h}\times M_{\rm v} \equiv 20 \times 20$ antennas. Moreover, from previous analysis to the ULA, we know that the $\Delta$ defines the range of the arrival angles. Thus, we can conclude that for the UPA the same behavior will be verified, in which when $\Delta_\varphi$ and $\Delta_\theta$ increase, the arrival signals become less correlated, and therefore increasing the capacity. We analyze the capacity for some fixed values of $\Delta$, being the elevation nominal angle defined as $0^o$. According to Fig. \ref{fig:22}.c, one can notice that the higher the value of $\Delta$, greater the capacity; however, the capacity only increases significantly if both $\Delta_\varphi$ and $\Delta_\theta$ increase simultaneously. As we know, when the BS is equipped with ULA, the elevation angle does not interfere on the capacity, unlike the UPA. Thus, the channel capacity using UPA arrangements is reduced compared to the ULA models. Lastly, in Fig. \ref{fig:22}.d the capacity is investigated according to the vertical and horizontal antenna spacing for different values of $\Delta_\varphi$ and $\Delta_\theta$. From this analysis, one can see that when both $\Delta_\varphi$ and $\Delta_\theta$ are simultaneously larger, the capacity is increased faster considering small vertical/horizontal antenna separation distances. The same oscillation characteristic observed in the One-ring channel model when the BS is equipped with ULA is also confirmed with UPA arrangements, where the capacity is also limited in $C\approx 350$ bit/s/Hz, due to the number of antennas is the same in both situations.

\subsubsection{Gaussian Model with UPA}
\begin{figure}[!htbp]
\centering
\includegraphics[width=0.495\textwidth]{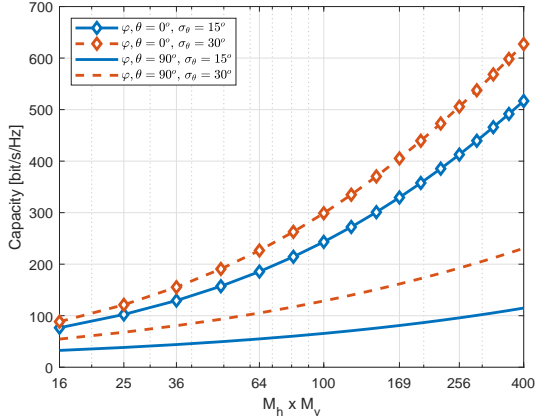}
\includegraphics[width=0.495\textwidth]{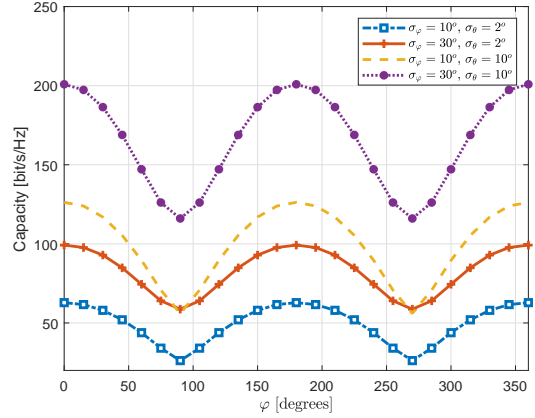}\\
{\bf a}) as a function of the number of antennas $M$ ($\sigma_\varphi = 30^o $) \hspace{8mm}  {\bf b}) azimuth nominal AoA ($M=100$, $\theta = 0^o$)  \\
\includegraphics[width=0.495\textwidth]{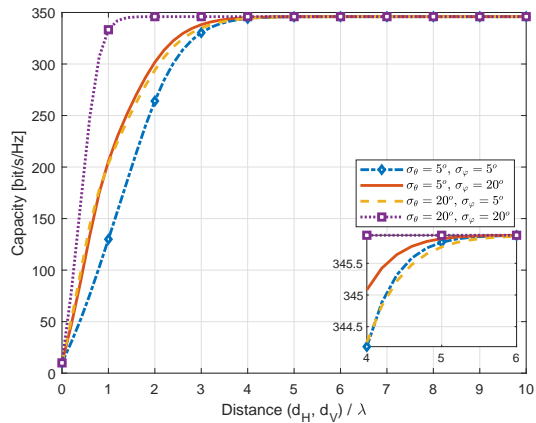}\\
{\bf c}) as a function of the spacing between the antennas ($\varphi,\, \theta= 0^o$ and $M=100$).
\label{fig:53}
\caption{Gaussian Model where BS is equipped with UPA. Capacity according to: a) increasing the number of antennas; b) azimuth nominal angle.}
\label{fig:37}
\end{figure}

The second 3D channel model for massive MIMO systems is the Gaussian model with UPA arrangements. As a numerical result, the channel capacity with increasing number of antennas is presented in Fig. \ref{fig:37} a. One can observe that when BS is equipped with both ULA and UPA, the Gaussian model results in a higher capacity than that provided by One-ring model, considering the same nominal values of $\Delta$ and $\sigma$, {\it i.e.}, the spread range is not equivalent in both models. Hence, with a uniform distribution, the spread is limited differently when compared with a Gaussian distribution, thus the angular range for the arrival signals in Gaussian model is larger, resulting in greater capacity gain. Moreover, the arrival angles are analized in Fig. \ref{fig:37} b. When compared with the One-ring UPA model, Fig. \ref{fig:22}.c, considering the same AoA and number of antennas, the Gaussian model presents higher channel capacity, since we know that this behavior is related to the antenna array geometry. For instance, the channel capacity predicted by the Gaussian UPA model when the nominal angles equal to zero and the respective standard deviations $\sigma_{\varphi}=30^o$ and $\sigma_{\theta}=10^o$, the capacity is  almost 200 bit/s/Hz while the One-ring UPA model predicts a capacity of $C\approx 105$ bit/s/Hz. The fact that the Gaussian model provides a larger angular arrival range than the One-ring model can be confirmed when we increase the $\Delta_{\theta}=30^o$ as depicted in the same Fig. \ref{fig:22}.c. The capacity in this case is $C\approx175$ bit/s/Hz, even smaller compared to that predicted by the Gaussian UPA model.

Finally, the behavior of the channel capacity based on Gaussian UPA model according to the spacing between the vertical/horizontal antennas is shown in Fig. \ref{fig:37}.c. As explained previously, the spacing between antennas is related to the degree of signal correlation and therefore the capacity increases with increasing antenna separation until saturate in $C\approx 350$ bits/s/Hz. In this analysis we simulate at first $\sigma_{\theta}=5^o$ and $\sigma_{\varphi}=20^o$ and in a second moment we analyze the capacity when  $\sigma_{\theta}=20^o$ and $\sigma_{\varphi}=5^o$.  Hence, the capacity results quite similar, evidencing the fact that the increasing in the spread range, measured by the standard deviation in the elevation $\sigma_\theta$ and the azimuth $\sigma_\varphi$ angles, implies in a consistent increment in thr channel capacity.

{\subsection{CBSM and GBSM Compared to Channel Measurements}}

{To corroborate the validity of the proposed stochastic channel models, in this section a performance analysis of the CSBM and GBSM models are compared to the  massive MIMO channel campaign measurement conducted  in \cite{hoydis2012}. The channel measurements were performed outdoors with a scalable virtual antenna array consisting of up to 112 ULA elements, representing a residential urban area. The deployed CBSM and GBSM channel parameter values are depicted in Table \ref{tab:4}.}
\newpage

\begin{table}[!htbp]
\centering
{\caption{Parameter values adopted in the numerical simulations for channel correlation analysis.}
\label{tab:4}
\begin{tabular}{|l|c|}
\hline
\multicolumn{1}{|c|}{\textbf{Parameters}}                      & \textbf{Values}                \\ \hline
\multicolumn{2}{|c|}{\bf Exponential Model}\\
	\hline
Path loss term                 & $\beta$ = 1     \\ 
\hline
Correlation index            &  $\rho=[0.8;0.85]$ \\  \hline
\multicolumn{2}{|c|}{\bf Exponential Model with Large Scale Fading}\\
	\hline
Path loss term                 & $\beta$ = 1     \\ 
\hline
Correlation index            &  $\rho=0.8$ \\ 
\hline
Shadowing -- standard deviation            & $\sigma_{\rm shad}=0$ dB   \\ 
\hline
Angle of arrival           & $\theta \sim \mathcal{U}(0,\,\theta_\text{max})$ degrees     \\ 
\hline
Maximum angle of arrival           & $\theta_\text{max}=[60;90;180]$ degrees     \\ 
\hline
\multicolumn{2}{|c|}{\bf One-ring ULA}\\
	\hline
Path loss term                 & $\beta$ = 1     \\ 
\hline
Angular interval of the angles of arrival           & $\Delta = 10$ degrees   \\ 
\hline
Angle of arrival           & $\varphi \sim \mathcal{U}(0,\,180)$ degrees   \\ 
\hline
Antenna spacing                 & $d_H$ = 0.5      \\ 
\hline
\multicolumn{2}{|c|}{\bf Gaussian ULA}\\
	\hline
Path loss term                 & $\beta$ = 1     \\ 
\hline
Shadowing -- standard deviation            & $\sigma_{\rm shad}$ = 0 dB   \\ 
\hline
Angular standard deviation                & $\sigma_{\varphi} = 10$ degrees      \\ 
\hline
Angle of arrival           & $\varphi \sim \mathcal{U}(0,\,180)$ degrees   \\ 
\hline
Antenna spacing                 & $d_H$ = 0.5      \\ 
\hline
\end{tabular}}
\end{table}

\begin{figure}[!htbp]
\centering
\includegraphics[width=0.49\textwidth]{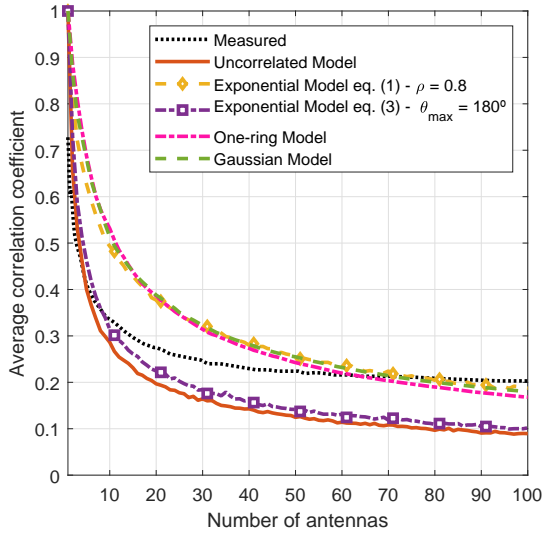}
\includegraphics[width=0.49\textwidth]{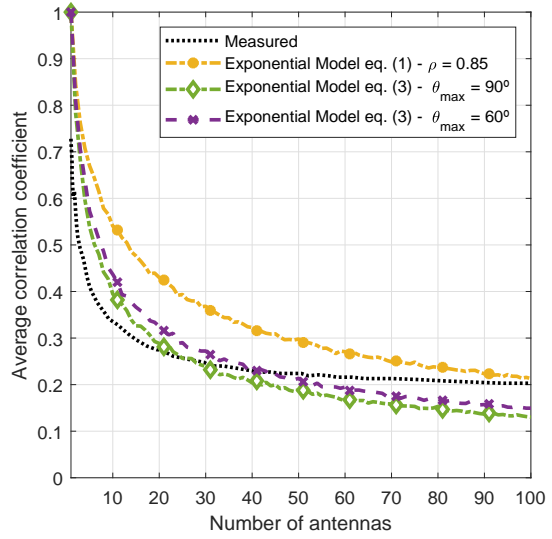}
{\bf a}) comparison between CBSM and GBSM; \hspace{20mm}{\bf b}) evaluation of other parameter values for CBSM.\hspace{5mm}
\caption{{Average correlation coefficient according to the ULA number of antennas : channel measurement vs CBSM and GBSM channel models with adjusted parameters.}}
\label{fig:23}  
\end{figure}

{For this analysis, we used the correlation coefficient between two channel vectors defined with:}
{
\begin{equation}
    \nu (M)=\frac{|\textbf{h}_{i,M}^H\textbf{h}_{j,M}|}{||\textbf{h}_{i,M}|| \hspace{0.1cm}||\textbf{h}_{j,M}||}
\end{equation}}
{where $\textbf{h}_{i,M}$ and $\textbf{h}_{j,M}$ are the channel vectors from the user i and user j, respectively. The correlation coefficients were evaluated according to the number of antennas $M$ for the models in which the BS is equipped with ULA. The results obtained for this simulation are depicted in Fig. \ref{fig:23}.}

{First, we can observe that both the measurement and the channel models show a decreasing correlation behavior  with the increase in the number of antennas. This result is due to the favorable propagation massive MIMO property, in which the orthogonality of the channels between users is attained when the number of antennas increases.
According to the data of the measured values, it is possible to observe that the correlation has a rapid decay when the number of antennas increases in the range [1; 10]. However, the correlation decays very slow, remains with few variations, when the number of antennas exceeds 50, {\it i.e.}, characteristic values of a massive MIMO system. In Fig. \ref{fig:23}a, if we divide the analysis according to the number of antennas, we see that the uncorrelated model and the exponential model with large-scale fading better capture the channel conditions for this outdoor channel measurement scenario. However, for a larger number of antennas, the GBSM models and the exponential model behave similarly to the measured values. Moreover, in Fig. \ref{fig:23}b, some channel model parameter values have been modified aiming at better fitting the measured channel. Hence, for the exponential model, we increased slightly the value of the correlation factor while for the exponential model with large-scale fading, the arrival angle was reduced from $\theta_{\max}=180^{\rm o}$ to $\theta_{\max}=90^{\rm o}$ and $60^{\rm o}$. We observed that the angle of arrival $\theta$ affects the correlation, in which for smaller values, the correlation coefficient increases. Such behavior is related to the confinement of the arrival signals, that is, if the $\theta$ assumes smaller values, the signals will be confined to a smaller angular interval. Therefore, given the results of each model, one can notice that this specific outdoor measurement environment presented a high degree of correlation, in which the parameters of each model could be adjusted in order to more accurately represent each environment.}

\subsection{XL-MIMO channel model}
In this section the validation of the analyzed XL-MIMO channel models is corroborated numerically. To evaluate the XL-MIMO scenario, it was used the SINR as a figure of merit. The adopted channel and system parameters values were described in Table \ref{tab:1}.

\begin{figure}[!htbp]
\centering
\includegraphics[width=0.44\textwidth]{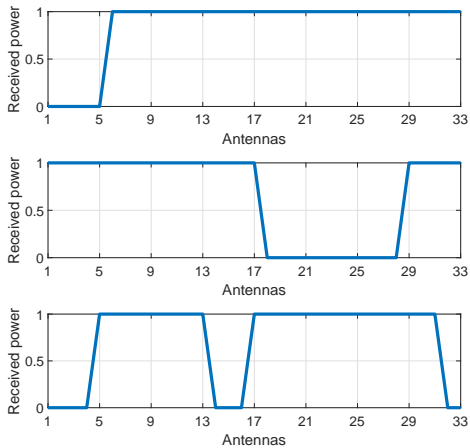}
\includegraphics[width=0.44\textwidth]{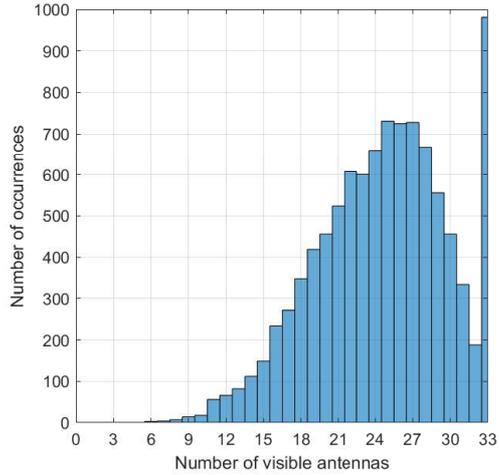}\\
{\bf a}) Three VR realizations \hspace{40mm}  {\bf b}) Number of active antennas distribution
\caption{VR generation algorithm considering ULA with $M=33$ antennas: a) Example of three VR, where the obstacles result in a received power equal to zero; b) histogram of the number of active antennas.}
\label{fig:62}
\end{figure}

First, we present an analysis of the algorithm to create the visibility region (VR). The algorithm was constructed so that obstructions usually involve more than one antenna, \textit{i.e.} the obstacle impacts on the received signal strength along the array antennas. Also, as distances between mobile terminal and BS antenna array are close, the likelihood of no obstacles is high. Fig. \ref{fig:62} depicts an example of  VR with obstructions generated randomly and the correspondent histogram indicating the number of visible (active) antennas. From the histogram, one can infer that it is highly probable that the $\approx 75\%$ of antennas ({\it i.e.}, 24 to 27 of 33 antennas) become visible or even 100\% of the antennas lie in the visibility region.

\begin{figure}[!htbp]
\centering
\includegraphics[width=0.46\textwidth]{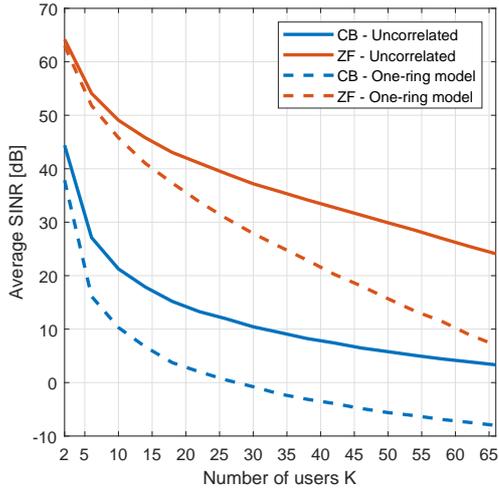}
\includegraphics[width=0.46\textwidth]{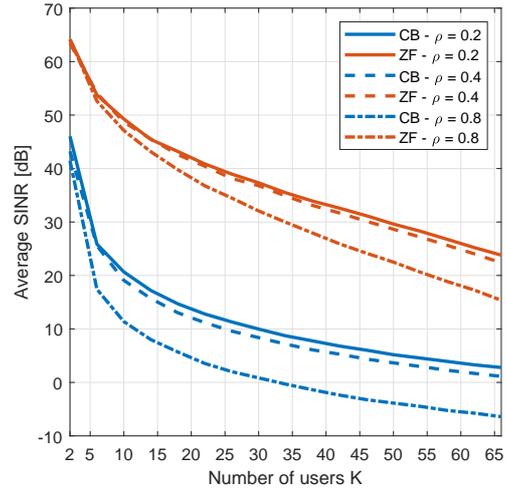} \\
{\bf a}) First scheme of clusters distribution \hspace{25mm}  {\bf b})  First scheme of clusters distribution \\
\includegraphics[width=0.46\textwidth]{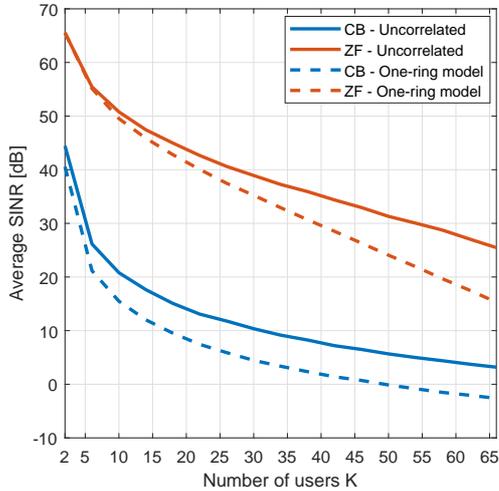}
\includegraphics[width=0.46\textwidth]{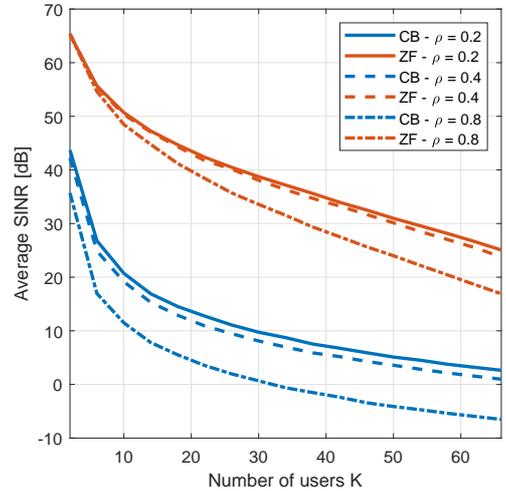} \\
{\bf c})  Second scheme of clusters distribution \hspace{25mm}  {\bf d}) Second scheme of clusters distribution  \hspace{5mm} 
\caption{Attainable SINR as a function of number of users ($K$) for  two schemes of clusters distribution: a)  Correlation matrix \textbf{R} is modelled by Uncorrelated and One-ring models; b) \textbf{R} is modelled by Exponential channel model. c)  Second scheme of clusters distribution: \textbf{R} is modeled by Uncorrelated and One-ring model; d) \textbf{R} is modeled by Exponential channel model.}
\label{fig:43}
\end{figure}

Besides, the SINR analysis considering both schemes of {\it cluster distribution} sketched in Fig. \ref{fig:41}, as well as both linear precoding conjugate beamforming (CB) and zero-forcing (ZF) and three correlated channel models, namely Uncorrelated fading, Exponential and One-ring have been carried out. Firstly, we compare the average SINR by user attainable with CB and ZF precoding for the first cluster distribution scheme from Fig. \ref{fig:41}.a, considering the correlation matrix as i.i.d. Rayleigh and the One-ring channel model, as depicted in Fig. \ref{fig:43}.a and \ref{fig:43}.b. The uncorrelated fading model implies that there is not correlation between antennas, unlike the One-ring model in which the channel correlation is implicit in the received signal via spread $\Delta$ parameter and between the antennas, according to the azimuth angle $\varphi$. Hence, the achievable SINR considering both CB and ZF precoding under uncorrelated channel model is higher than that attainable under One-ring channel model and the respective linear precoding for any number of users. However, when the number of users increases, the SINR  is reduced, considering both linear precoders, while the SINR values difference provided by the uncorrelated {\it vs} One-ring channel models
increases. Indeed,  the behaviour of the SINR is similar considering both linear precoders, but CB and ZF linear precoders implement different rules. The  CB precoding focus is on increasing the power of the desired user, while, the ZF focus on decreasing or virtually reduce to zero the power interference. Thus, one can observe that the ZF precoding presents higher SINR's than the CB; however, it is known that there is a maximum number of users in which the ZF can eliminate completely the interference, and when that value is exceeded, the performance of the ZF precoding is reduced. 

Furthermore, as the Exponential channel model depends of the correlation factor, the analysis was developed separately and presented in Fig. \ref{fig:43}.b. One can observe that the reduction in the SINR is impacted by the system loading $\frac{K}{M}$, as well as by the increase of the correlation factor $\rho$. As discussed previously,  in the Fig. \ref{fig:8}.b), the channel capacity is reduced substantially when the channel correlation factor approach to 1, being this impact more pronounced when the correlation factor is $\rho>0.6$. Hence,  the same characteristic is corroborated for the SINR analysis in Fig. \ref{fig:43} when the $\rho=0.8$. 

Considering the second cluster distribution scheme where all users are aligned and parallel to the ULA arrangement, as sketched in Fig. \ref{fig:41}.b, we calculate the average SINR performance when the One-ring and uncorrelated fading channel models are considered; the results are depicted in Fig. \ref{fig:43}.c and \ref{fig:43}.d. The behavior of the SINR is same for the uncorrelated if compared to the first scheme. However, for the One-ring model the attainable SINR is higher than that achieved under the first cluster distribution scheme. The reason of the first scheme presented worst performance is due to distribution region of the clusters are in locations where the received signal is perpendicular to the antenna array, increasing the correlation between the antennas (see Fig. \ref{fig:31}). Finally, in the Fig. \ref{fig:43}.d, the Exponential channel model under second cluster distribution scheme is analyzed. The same SINR gain can be observed comparing Fig. \ref{fig:43}.b and \ref{fig:43}.d for the Exponential channel model. The results demonstrate that the attainable average SINR provided by each linear precoding depends strongly on the signal propagation environment. Nevertheless, only the One-ring model properly describes the propagation characteristics because it is a geometric-based channel model. Thus, the Uncorrelated and Exponential channel models do not show SINR performance differences because they are correlation-based models, where for the Exponential model, the impact of environmental propagation on the degree of correlation is indirectly considered by a correlation factor. From our numerical simulated results, one can infer that for the second cluster distribution scheme using the One-ring model, an Exponential model with a correlation factor of $\rho=0.8$ have resulted in a similar SINR behavior, indicating a high degree of correlation for this distribution on the One-ring model.

\section{Conclusions}\label{sec:concl}
Channel models for massive MIMO and extreme large (XL) MIMO systems addressing relevant features implicit in such rich and innovative communication system scenarios are not completely available in the literature. Recent works have investigated stochastic channel models, such as correlated-based (CBSM) and geometric-based (GBSM) stochastic models due to the possibility of implementing massive MIMO channel models with low-computational complexity and considering physical characteristics of signal propagation. This paper focused on an extensive analysis and comparison of these channel models in both stationary and non-stationary scenarios aiming at M-MIMO and XL-MIMO applications. We have discussed how the characteristics such as shadowing, angle of arrival, and angular spread can emulate the channel correlation degree implicitly. Among 2D and 3D GBSMs, the 2D channel model presents higher capacity gain due to the possibility of the received signals between the antennas result uncorrelated at any elevation angle. However, the model with the highest capacity gain is the CBSM. Although the CBSM model presented higher capacity gain, for the XL-MIMO applications, both channel models when compared considering two representative cluster distribution schemes, only the GBSM model has resulted SINR performance variations. Thus, the GBSM model proved to be more suitable for XL-MIMO channel modelling, since the propagation parameters are considered in the model.


\section*{Acknowledgments}
This work was supported in part by the National Council for Scientific and Technological Development (CNPq) of Brazil under Grants 310681/2019-7, and in part by CAPES - Coordena\c{c}\~ao de Aperfei\c{c}oamento de Pessoal de N\'ivel Superior, Brazil (scholarship, Finance Code 001),  and by the Londrina State University - Paran\'a State Government (UEL).

\end{document}